\documentclass[conference]{IEEEtran}

\usepackage{acronym}
\usepackage{amsmath}
\usepackage{array}
\usepackage{booktabs}
\usepackage{censor}
\usepackage{cite}
\usepackage{colortbl}
\usepackage{dblfloatfix}
\usepackage{float}
\usepackage{graphicx}
\usepackage{makecell}
\usepackage{multirow}
\usepackage{subcaption}
\usepackage{tikz}
\usepackage{pgfplots}
\usepackage{pgfplotstable}
\usepackage[most]{tcolorbox}
\usepackage{url}
\usepackage{xcolor}
\usepackage{xspace}
\usepackage{booktabs}

\pgfplotsset{compat=1.18}
\usetikzlibrary{pgfplots.groupplots}
\usepgfplotslibrary{groupplots,polar,statistics}
\graphicspath{{Images/}}
\definecolor{customblue}{RGB}{0,102,204}
\definecolor{customdarkblue}{RGB}{0,35,150}
\definecolor{customred}{RGB}{180,40,40}
\definecolor{customgrey}{RGB}{90,90,90}
\definecolor{customgreen}{RGB}{0,140,75}

\acrodef{AI}{artificial intelligence}
\acrodef{CTI}{Cyber threat intelligence}
\acrodef{DL}{deep learning}
\acrodef{LLM}{large language model}
\acrodef{ML}{machine learning}
\acrodef{NER}{named entity recognition}
\acrodef{NLP}{natural language processing}
\acrodef{OSINT}{Open source intelligence}
\acrodef{IoCs}{indicators of compromise}
\acrodef{IoC}{indicator of compromise}
\acrodef{TTPs}{tactics, techniques, and procedures}
\acrodef{TTP}{tactic, technique, and procedure}

\newcommand{\SysName}{DarkBot\xspace}
\newcommand{\bl}[1]{\textcolor{customdarkblue}{#1}}

\newcommand{\mypari}[1]{\smallskip\noindent\textit{#1.}\xspace}
\newcommand{\mypar}[1]{\smallskip\noindent\textit{\textbf{#1.}}\xspace}
\newcommand{\myparnotitle}{\smallskip\noindent}
\newcommand{\myparq}[1]{\smallskip\noindent\textit{\textbf{#1}}\xspace}
\usepackage{todonotes}

\tcbset{
  systurn/.style={
    enhanced,colback=white,boxrule=0pt,
    borderline west={1.5pt}{0pt}{customred},
    borderline north={0.3pt}{0pt}{black!30},
    borderline south={0.3pt}{0pt}{black!30},
    borderline east={0.3pt}{0pt}{black!30},
    sharp corners,left=4pt,right=4pt,top=2pt,bottom=2pt,
    fontupper=\small\ttfamily,
  },
  userturn/.style={
    enhanced,colback=white,boxrule=0pt,
    borderline west={1.5pt}{0pt}{customgreen},
    borderline north={0.3pt}{0pt}{black!30},
    borderline south={0.3pt}{0pt}{black!30},
    borderline east={0.3pt}{0pt}{black!30},
    sharp corners,left=4pt,right=4pt,top=2pt,bottom=2pt,
    fontupper=\small\ttfamily,
  },
}

\newcommand{\myexamplebox}[5][t]{%
  \begin{figure}[#1]
  \small
  \begin{tcolorbox}[
    enhanced,colback=white,colframe=black!50,boxrule=0.4pt,
    sharp corners,left=4pt,right=4pt,top=3pt,bottom=3pt]
  \begin{tcolorbox}[
    colback=gray!5,colframe=gray!90,boxrule=0.4pt,sharp corners,
    title=\textbf{Thread Initial Post},fonttitle=\bfseries,
    left=3pt,right=3pt,top=2pt,bottom=2pt]
  #3
  \end{tcolorbox}
  \smallskip
  #4
  \end{tcolorbox}
  \caption{#2}
  \label{fig:#5}
  \end{figure}
}

\newtcolorbox{takeaway}{
  enhanced,
  colback=gray!5,
  colframe=black,
  boxrule=0.6pt,
  sharp corners,
  boxsep=1.5pt,
  left=1pt,right=1pt,top=1pt,bottom=1pt,
  fonttitle=\bfseries,
}

\begin{document}

\title{Towards Automated Cyber Threat Intelligence Elicitation in Underground Forums}

\author{
\IEEEauthorblockN{
Lorenzo Bossi\IEEEauthorrefmark{1},
Federico Saccani\IEEEauthorrefmark{1},
Francesco Panebianco\IEEEauthorrefmark{1},
Antonio Maci\IEEEauthorrefmark{2}\IEEEauthorrefmark{1},\\
Stefano Zanero\IEEEauthorrefmark{1},
Stefano Longari\IEEEauthorrefmark{1},
Michele Carminati\IEEEauthorrefmark{1}
}

\IEEEauthorblockA{
\IEEEauthorrefmark{1}Politecnico di Milano
}

\IEEEauthorblockA{
\IEEEauthorrefmark{2}BV TECH S.p.A.
}

\IEEEauthorblockA{
\texttt{\{lorenzo.bossi, francesco.panebianco, antonio.maci, stefano.zanero,}\\
\texttt{stefano.longari, michele.carminati\}@polimi.it}\\
\texttt{federico.saccani@mail.polimi.it}
}
}

\maketitle

\begin{abstract}
Cyber threat intelligence from underground forums has traditionally relied on passive monitoring. However, as users have become more aware of large-scale data collection, valuable intelligence has become increasingly rare in open forums, often migrating instead to private or harder-to-reach spaces, making passive approaches inadequate. 
Building on the intuition that relevant information can be obtained through active elicitation, this paper presents \SysName, to the best of our knowledge, the first multi-agent LLM-based system for active CTI elicitation in underground forums. \SysName decomposes the interaction task across eleven specialized agents organized into three functional blocks: engagement gating for relevance and safety filtering, context-aware question generation driven by MITRE ATT\&CK tactics, and linguistic style adaptation to better align with real forum users. 
In a controlled evaluation across 100 CrimeBB conversations, the system recovered 72.8\% of the validated MITRE ATT\&CK techniques present in the original discussions by observing only the initial post at the start of each interaction, and it consistently outperformed a monolithic baseline. The proposed layered safety design contained all injected jailbreak attempts at the pipeline level. These results were further supported by real-world experiments: in a prospective matched deployment, threads assigned to \SysName accumulated an average of 3.85 more CTI entities than their controls over seven days, and across 104 live forum conversations, the system elicited CTI-relevant disclosures without observed account suspensions, moderator interventions, or explicit accusations of automated participation.
\end{abstract}


\section{Introduction}
\ac{CTI} is the process of collecting, analyzing, and interpreting cyber threat data to support proactive security decisions. 
Rather than focusing only on raw artifacts such as \ac{IoCs}, it turns heterogeneous and often unstructured information into contextualized actionable insights (e.g., attack \ac{TTPs}, motivations, and likely targets) aligned with an organization's specific risk profile \cite{AINSLIE2023103352}.
In contemporary threat landscapes, a significant portion of \ac{CTI} is derived from online communities, such as public forums and underground platforms, which play a central role in this ecosystem, as discussions within them often anticipate the emergence of vulnerabilities, data leaks, or attack campaigns~\cite{you_might_have_known_it_earlier_polimi}.
\ac{CTI} data collection practices are largely based on passive strategies, including automated crawling and monitoring of publicly accessible sources.
However, the structure of these ecosystems is evolving, as recent studies suggest a gradual shift towards more private or hidden platforms, with users becoming more aware of large-scale data collection, reducing the visibility of operationally relevant information in open forums \cite{you_might_have_known_it_earlier_polimi}. 
Moreover, online discussions are inherently interactive; relevant information often emerges through dialog, clarification, and targeted questioning rather than appearing spontaneously.
These characteristics highlight a key limitation of purely passive collection approaches, which may fail to surface valuable insights that require active engagement.

In parallel, recent advances in generative \ac{AI} have fostered growing interest in the use of \acp{LLM} within \ac{CTI}~\cite{10198233}. 
These models are increasingly adopted as components of automated \ac{CTI} pipelines, where they assist both downstream analysis and earlier stages, such as data collection, by processing and interpreting large volumes of unstructured content \cite{Balasubramanian2025}. 
More recent approaches further investigate multi-agent \ac{LLM}-based systems that enable distributed reasoning and task decomposition, improving the robustness and depth of extracted intelligence through coordinated interactions among specialized agents \cite{10679480,mad-cti}.

Building on these insights, we explore a novel paradigm for \ac{CTI} data collection based on controlled interaction with online communities. Specifically, we investigate whether carefully designed conversational agents can stimulate discussions in publicly accessible environments to generate relevant threat intelligence.
Unlike existing multi-agent CTI systems, which primarily focus on offline analysis of previously collected data (e.g., relevance classification or static content triage) {\cite{mad-cti}}, the proposed approach targets active and iterative information elicitation through direct interaction with real users. We move from passive analysis to credible undercover participation in underground online discussions, where the system engages as a regular user to extract CTI-relevant insights.

Consequently, we propose an \ac{LLM}-based multi-agent system, namely~\SysName, that operates during the data-collection phase. The system coordinates multiple specialized agents for contextual analysis, linguistic adaptation, and question generation, enabling targeted, context-aware information elicitation. 
Designed to operate in adversarial and dynamic environments such as underground forums,~\SysName must ensure safe engagement, maintain contextual awareness, and produce linguistically credible interactions. To this end, it follows an iterative workflow: given an initial thread,~\SysName evaluates whether engagement is appropriate, generates context-consistent questions, and continuously analyzes incoming replies to decide whether to continue or terminate the interaction. Rather than relying on a single monolithic model, the architecture decomposes the overall task into subtasks handled by multiple agents organized into three main functional blocks. The engagement gating block addresses safety and relevance by filtering prohibited content and detecting adversarial inputs. Second, the context understanding and question generation block focuses on thread analysis, elicitation planning, and target question generation. Lastly, the linguistic style and credibility block ensure alignment with the communication style of real users, improving the ability of~\SysName to blend into the target environment.
Our evaluation shows that, across 100 simulated CrimeBB conversations, ~\SysName recovers 72.8\% of the MITRE ATT\&CK techniques present in the original conversation using targeted questions, starting only from the initial post, while ablation studies confirm that multi-agent decomposition consistently outperforms a monolithic baseline. 
The layered safety architecture contained 100\% of the tested jailbreak attempts at the end-to-end pipeline level, despite imperfect individual components. In isolation, the \textsc{DefensiveAgent} detected 92.8\% of jailbreak prompts, while the final \textsc{FilterAgent} blocked 85.6\% of slang-rewritten policy-violating messages in a separate worst-case evaluation, with undetected questions being borderline content. 
In a prospective experiment with 20 matched topic pairs, threads assigned to \SysName accumulated on average 3.85 more CTI entities 
compared to their control.
Deployed across 104 live conversations and 213 published messages on surface- and dark-web forums,~\SysName elicits CTI that would not have surfaced without active probing, including details of active data breaches and newly published attack techniques already being tested in the wild, without ever triggering bans, moderator actions, or AI accusations.

The contributions of this paper are the following:

\begin{itemize}
    \item We introduce \SysName, to the best of our knowledge, the first multi-agent LLM-based system for active CTI elicitation in public underground forums. The system decomposes the interaction process into eleven specialized agents responsible for engagement gating, ATT\&CK-guided elicitation planning, question generation, adversarial-input handling, and community-specific linguistic adaptation.
    \item We design and evaluate a layered safety architecture for operation in adversarial forum environments. The architecture combines relevance filtering, reply analysis, jailbreak detection, and post-generation filtering, detects jailbreak attempts at the pipeline level, and blocks most policy-violating outgoing messages.
    \item We evaluate active participation in real forum environments through both a prospective matched experiment and deployment in 104 live conversations. The system elicited CTI-relevant disclosures without account suspensions, moderator interventions, or explicit accusations of automated participation during the study period.
\end{itemize}

\section{Background}

\ac{OSINT} and \ac{CTI} are central to modern cybersecurity. In this context, \ac{OSINT} refers to the collection and analysis of information from publicly accessible or semi-open sources, including websites, repositories, vulnerability databases, and underground forums \cite{11143131,11015496,11129557,sok_cti}. 
\ac{CTI} builds on this information to produce structured knowledge about threat actors, techniques, vulnerabilities, and malicious infrastructure for defensive decision-making \cite{10117505}. A key reference in this process is the MITRE ATT\&CK framework~\cite{mitre,mitre_attack}, which organizes adversarial behavior into \ac{TTPs} and helps guide both threat collection and interpretation \cite{10.1145/3687300,10539631}. 
Recent work has expanded \ac{OSINT} and \ac{CTI} through large-scale data analysis and \ac{AI}-based methods.
In particular, \acp{LLM} are being progressively incorporated into \ac{CTI} extraction pipelines because of their ability to support multiple steps of the process, including knowledge assessment, vulnerability analysis, and threat actor attribution, as highlighted by benchmark-oriented evaluations such as CTIBench \cite{NEURIPS2024_5acd3c62}. In more specialized settings, they have been used for report summarization and data augmentation to improve downstream classification, e.g., for extracting MITRE ATT\&CK techniques from unstructured \ac{CTI} reports \cite{10.1145/3701716.3715469}.
More recent systems also use \acp{LLM}, combining fine-tuning or in-context learning for entity and relation extraction with retrieval-augmented reasoning modules to support attack chain inference and question answering over structured \ac{CTI} knowledge \cite{Yang2026}.
Techniques such as zero-shot and few-shot prompting enable such models to perform tasks ranging from question answering and summarization to reasoning and decision support by embedding task instructions and examples directly within the prompt \cite{NEURIPS2020_1457c0d6,10.5555/3600270.3602070}. 
Recently, research has explored prompt chaining and multi-agent \ac{LLM} architectures to address more complex workflows. In these approaches, multiple prompts or specialized agents interact sequentially, with the output of one model invocation serving as the input for subsequent reasoning steps. Prompt chaining allows tasks to be decomposed into smaller subtasks, enabling more structured reasoning. Similarly, multi-agent frameworks employ multiple LLM-based agents with specialized roles that collaborate to solve complex problems through iterative interaction and information exchange \cite{autogen}. 
These approaches leverage the compositional capabilities of \acp{LLM}~\cite{ahuja2024provable,xularge} to orchestrate more sophisticated reasoning pipelines than those achievable with a single prompt invocation.

\section{Related Work}

Previous work has explored chatbots for investigative and forensic purposes, mainly in three distinct directions. First, conceptual work~\cite{10.1145/3469595.3469598} has proposed chatbots as intelligence-gathering agents for evidence collection on the dark web, highlighting their potential for continuous operation, rapid interaction, and support for law-enforcement investigations, but without presenting a fully implemented or evaluated system. Second, works such as ScamChatBot \cite{acharya2024scamchatbotendtoendanalysisfake} have shown that \acp{LLM} can be used to actively engage malicious actors at scale and uncover scam tactics, coordination patterns, and payment infrastructures.
Similarly, Yao et al.~\cite{yao2025imitation} deploy an LLM to engage with illicit scammers on Telegram, demonstrating conversational infiltration, but using a single-agent design and targeting transactional indicators (e.g., payment methods). However, these systems focus on social media scam ecosystems rather than on recurring engagement for CTI elicitation.
Third, other research \cite{qonita} investigated chatbot-based support for digital forensics, for example, by assisting investigators in analyzing logs, emails, and image metadata, but these systems operate primarily on already-collected evidence rather than through live, adversarial interaction with forum users. More broadly, LLMs have also been shown to pose a privacy threat to ordinary users, as adaptive psychological profiling and communication strategies can be weaponized to stealthily elicit personal information~\cite{zhang2025power}. Taken together, these studies suggest that conversational agents can support cyber investigations both passively and actively, yet they leave open the question of how such systems should be designed to elicit sustained, credible, and goal-directed information in underground communities.

In parallel to the aforementioned strategies, multi-agent \ac{LLM} frameworks have been proposed to decompose other complex security tasks into specialized roles (e.g., passive analysis of forum conversations~\cite{mad-cti}, classification, translation, planning), often improving performance on complex pipelines \cite{11418024}. 

\mypar{Positioning of this work}
To the best of our knowledge, no prior work has investigated multi-agent \ac{LLM} systems that actively engage in recurring, interactive conversations with human users on underground forums to elicit \ac{CTI}-related information.
Thus, we design and evaluate an interactive multi-agent \ac{LLM}-based system for \ac{CTI}-oriented information elicitation in public underground forums. 
We investigate whether role decomposition across multiple agents can improve the effectiveness, robustness, and control of interactive \ac{CTI} elicitation in adversarial conversational settings.

\section{Motivation and Problem Statement}
\label{sec:motivation}

Public underground forums have historically been a valuable source of \ac{CTI} \cite{11129557}. 
However, in recent years, the quantity and quality of information available in these public spaces have steadily decreased \cite{you_might_have_known_it_earlier_polimi}, as threat actors have progressively migrated their most relevant discussions to less public, harder-to-reach spaces.
Meanwhile, the rapid advancement of large language models opens new opportunities for automated interactive collection methods, potentially enabling agents to stimulate discussions and surface relevant knowledge from private channels into public spaces, where it can be analyzed for proactive defense.
Formally, starting from an initial state of a forum thread, comprising the original post and any replies prior to the system's intervention, a passive monitoring approach can extract a set of CTI items (e.g., \ac{TTPs} and \ac{IoCs}). The system then engages in an iterative, turn-based interaction: at each turn, it observes the current thread state and decides either to publish a follow-up message or to terminate. After the interaction, the final thread state contains the original content plus all bot-elicited exchanges. From the final state, discarding the bot's own messages and retaining only human 
contributions, one can extract a set of CTI items. The objective is to maximize the \textit{CTI delta}, that is, the intelligence present in the human contributions of the full conversation that was not already present in the thread before the system intervened. A system is considered more effective the larger the CTI delta, meaning that active engagement surfaced intelligence that passive observation of the initial state alone would have missed.
In practice, however, the problem imposes three additional constraints: (1) The system must generate messages that are not easily distinguishable from those of legitimate users, as detection would compromise the system's ability to operate, (2) The system must be capable of resisting adversarial manipulation as it operates in an adversarial setting, (3) The system must bound the frequency and number of turns per thread and recognize when to terminate, as overly insistent interactions raise suspicion and increases detection risk.
These practical constraints raise three fundamental challenges in the design of an automated system capable of solving this problem.

\mypar{Adversarial environment} The system operates in an environment where human users may be actively suspicious, adversarial, or technically sophisticated. Standard LLM safety mechanisms, designed for cooperative settings, are insufficient. They may fail to detect manipulation attempts that exploit the conversational context, and they tend toward over-refusal:  a model that rejects any discussion of malware, exploits, or illegal activity cannot function in underground forums. 

\mypar{Sparse and implicit signals} CTI in underground forums is rarely stated explicitly as it is in more structured sources (e.g., threat reports and vulnerability databases). Relevant intelligence is embedded in fragmented and informal language. The system needs to understand the context and strategically formulate questions that maximize the likelihood of obtaining CTI-relevant information while minimizing the risk of detection. Standard LLMs, however, tend to produce generic conversational replies rather than targeted elicitation questions.

\mypar {Linguistic adaptation}
General-purpose LLMs, trained with alignment procedures that suppress informal or adversarial language, produce outputs that are stylistically out of place in underground forum settings~\cite{paech2025antislop,donmez2025ai,durandard2025llms}.

\myparnotitle
\begin{takeaway}
\textbf{Research Goals.}
The goal of this work is to investigate whether interactive \ac{LLM}-based agents can effectively be deployed to shift from passive CTI monitoring to active elicitation through targeted participation in underground forum discussions. 
To do so, we first address how to design a system that allows for safe yet credible operations in adversarial online environments, and decompose the elicitation task across multiple specialized agents.

\end{takeaway}

\section{Proposed Methodology}
\label{sec:approach}

\begin{figure*}[t]
  \centering
  \includegraphics[width=0.8\linewidth]{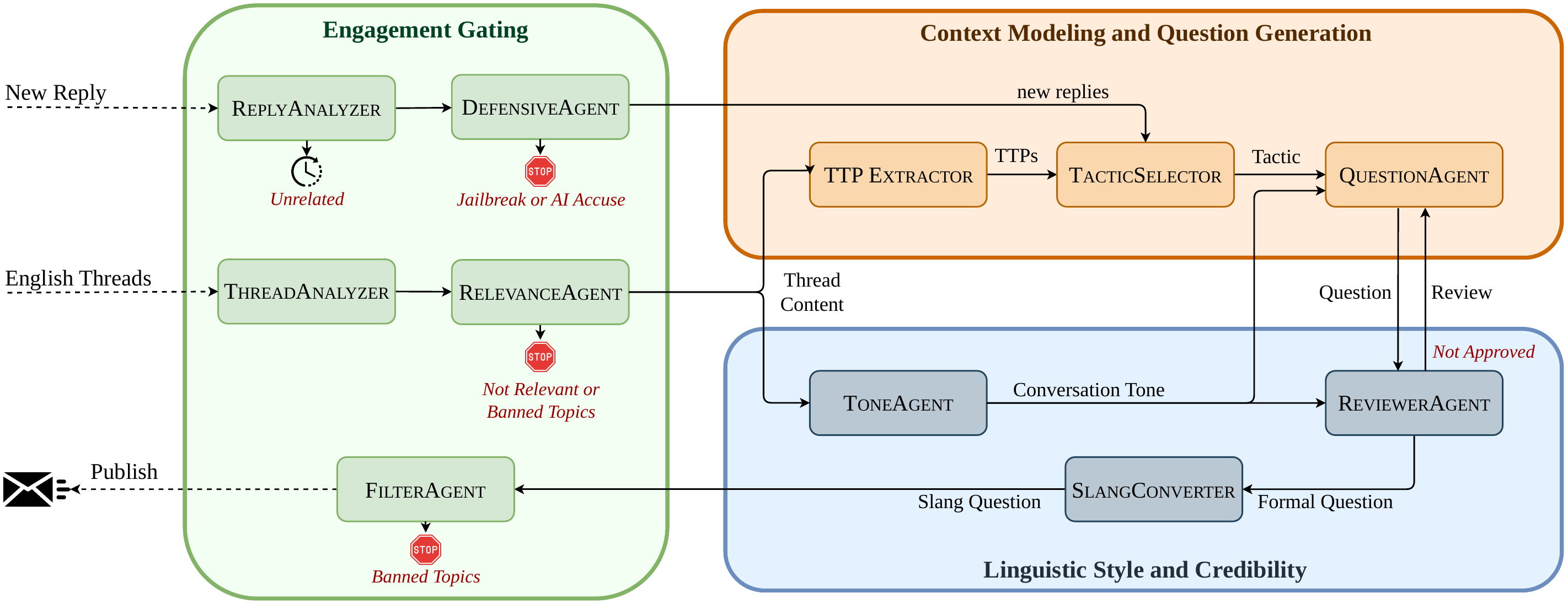}
  \caption{\SysName Architecture}
  \label{fig:system_architecture}
\end{figure*}

We design \SysName, an \ac{LLM}-based multi-agent system that participates in public underground forums acting as a regular user to elicit \ac{CTI}-relevant information through targeted questions. The system operates in iterative rounds: given an initial thread context, it determines whether engagement is allowed and worthwhile, it generates candidate questions consistent with the ongoing discussion, and iteratively inspects new replies to continue or terminate the interaction.
Figure~\ref{fig:system_architecture} summarizes the proposed pipeline, which is composed of 11 LLM agents who collaborate, breaking down a complex task into subtasks. Instead of a single prompt handling relevance, safety, discourse style, planning, and question generation, the system assigns these subtasks to separate agents and combines their outputs using explicit decision logic.

\subsection{Design Principles}

The CTI elicitation task presents multiple distinct challenges (\S\ref{sec:motivation}) that motivate the design principles of \SysName. As the problem is complex and composed of multiple parallel challenges, we adopt a multi-agent architecture as a foundational design choice, as it allows decomposing the problem into specialized agents, and has demonstrated consistent gains in complex security tasks~\cite{pentestgpt,mad-cti}. 

To address the~\textit{adversarial environment}, we introduce a dedicated engagement gating stage that restricts interaction to CTI-relevant threads to avoid unnecessary interactions, detects and terminates conversations involving jailbreak attempts or accusations of being a bot, and enforces compliance with our ethical guidelines (\S~\ref{sec:app_ethics}).

To address the \textit{sparse and implicit signals} challenge, we adopt a tactic-driven question generation mechanism organized around MITRE ATT\&CK tactics rather than techniques. 
Tactics correspond to high-level attacker goals (e.g., Initial Access and Privilege Escalation) that align naturally with the focus of most forum threads, whereas individual techniques often describe implementation variants that do not emerge as distinct conversational topics. Framing questions at the tactic level maximizes response probability from forum participants, while still allowing the bot to consider specific techniques that arise organically. Furthermore, with 14 tactics versus over 200 techniques, tactic-level tracking is a tractable problem for an LLM, whereas discriminating across hundreds of similar technique entries is considerably more complex, as confirmed by results in Section~\ref{sec:exp_ttp_extraction}.

To address the \textit{linguistic adaptation} challenge, we consider two dimensions important for credible participation. First, matching the thread's conversational tone (e.g., informal, aggressive) helps generate messages that blend naturally into the discussion and may improve elicitation, as perceived similarity favors information sharing~\cite{giles1991,cialdini1993}. Second, underground forums use community-specific slang and terminology that aligned models are likely to suppress.

\subsection{Functional Blocks}

We organize the pipeline dividing the underlying agents into three functional blocks addressing the identified challenges. Table~\ref{tab:agents} shows their goals, inputs, and outputs.

\mypar{Engagement Gating}
\label{sec:gating}
Before generating any message, the system applies a strict engagement gating stage to ensure it interacts only with threads that are \ac{CTI}-relevant and do not involve prohibited or high-risk categories. Concretely, the \textsc{ThreadAnalyzer} and \textsc{RelevanceAgent} follow a pattern proposed by Shah et al.~\cite{mad-cti}, where an analysis agent produces a structured thread report (what is being discussed, what technical artifacts are mentioned, what information is missing) and a gating agent then filters out threads that are irrelevant or fall under prohibited content categories  (e.g., violence, narcotics, sexually explicit content). If the thread violates constraints, the pipeline terminates without posting or saving any data.
The same pattern is applied in the analysis of subsequent replies using the \textsc{ReplyAnalyzer} and the \textsc{DefensiveAgent} to identify potential deviations towards restricted topics, jailbreak attempts, or accusations that the user is an AI bot. The \textsc{ReplyAnalyzer} detects if the new replies are related to previous bot questions or if they contain unrelated content. In the latter case, the bot waits for new replies.
A reply can be both topically relevant and adversarially manipulative, or topically irrelevant and entirely benign. Bundling these decisions into a single agent risks mixing the two, leading to underdetection of manipulation attempts embedded in otherwise relevant content. This risk is well-documented in the jailbreak literature~\cite{in_the_wild, do_anything_now}, which shows that adversarial prompts frequently exploit conversational context to bypass safety filters that reason only about content categories.
Furthermore, topic restrictions are also enforced post-generation of the system's response by the \textsc{FilterAgent}. Any candidate response that violates the topic policy is discarded and not published.

The prompt-based gating approach employed for both new threads, received replies, and generated messages was preferred over repurposing off-the-shelf safety classifiers such as LlamaGuard3~\cite{dubey2024llama3herdmodels}, which, being calibrated for general-purpose content moderation, classifies most hacking-related content as unsafe, suppressing the CTI-relevant discussions the system is designed to engage with (Appx~\ref{sec:app_llamaguard}).

\mypar{Context Modeling and Question Generation}
\label{sec:context_planning_generation}
The result is a structured list of attack behaviors and MITRE tactics, which serves as the foundation for subsequent reasoning. For the design, we referred to the project by Yuancheng Liu~\cite{llm_mitre}. Note that we adapted the original project for better integration into our pipeline, using the original prompts with minor changes to ensure greater consistency with the interaction of our agentic part. 
Based on this structured representation, the \textsc{TacticSelector} determines which MITRE tactic should guide the next interaction. The agent maintains a dynamic tactic map covering all fourteen MITRE tactics, which can be updated after each interaction. Each tactic is associated with a score (Eq.~\ref{eq:tactic_score}) from a prompt-based heuristic inspired by literature, and validated empirically (\S\ref{sec:exp_extraction}, \S\ref{sec:exp_real_forums}). 
\begin{equation}
\label{eq:tactic_score}
\footnotesize
  0.30 \cdot \mathcal{E}_{\text{str}} +
  0.25 \cdot \mathcal{R}_{\text{prob}} +
  0.15 \cdot \mathcal{I}_{\text{value}} +
  0.10 \cdot \mathcal{C}_{\text{flow}} +
  0.20 \cdot \mathcal{M}_{\text{appl}}
\end{equation}

The heuristic scores each candidate tactic along five dimensions: (1) \textbf{Evidence Strength} ($\mathcal{E}_{\text{str}}$), measuring concrete CTI evidence such as explicit technique mentions, behavioral patterns, and tool references~\cite{zibak2022threat}; (2) \textbf{Response Probability} ($\mathcal{R}_{\text{prob}}$), estimating the likelihood of a meaningful reply from prior response length and depth, topic fit, and timing, since even informative tactics are ineffective if they cause disengagement~\cite{oleszkiewicz2014scharff}; (3) \textbf{Intelligence Value} ($\mathcal{I}_{\text{value}}$), rewarding the expected novelty and actionability of elicited information~\cite{schlette2020measuring,dimitriadis2025evacti}; (4) \textbf{Conversational Flow} ($\mathcal{C}_{\text{flow}}$), penalizing abrupt topic shifts that disrupt coherence and elicitation effectiveness~\cite{oleszkiewicz2014scharff,nunan2022impact}; and (5) \textbf{MITRE ATT\&CK Applicability} ($\mathcal{M}_{\text{appl}}$), rewarding tactics that map concretely to ATT\&CK while promoting diversity across the kill chain~\cite{al2024mitre,li2024automated}.

\myparnotitle
The agent selects the highest-scoring tactic and terminates the pipeline if none is sufficiently effective. The \textsc{QuestionAgent} then combines the selected tactic with behaviors identified by the \textsc{TTP Extractor} to generate follow-up questions that deepen the tactical dimension, while avoiding procedural instructions or encouraging specific actions.

\mypar{Linguistic Style and Credibility}
Before generating a question, the \textsc{ToneAgent} analyzes the conversation to detect its dominant tone and the frequency of specific stylistic patterns (e.g., emojis and extreme punctuation). Available emotion classifiers~\cite{hartmann2022emotionenglish} capture affective categories but do not explicitly identify the stylistic features used, motivating a dedicated prompt-based component. This information guides the \textsc{QuestionAgent} to formulate a message that is consistent with the conversational context while preserving its original semantic intent. We keep the \textsc{QuestionAgent} as an independent component from the \textsc{TacticSelector} so that tactical planning and question generation can be handled independently; ablation results (\S\ref{sec:exp_design}) confirm the benefit of this decomposition.
After the \textsc{QuestionAgent} generates the candidate questions, they are evaluated by a dedicated \textsc{ReviewerAgent} following a reflection pattern ~\cite{microsoft_autogen_reflection_pattern, shinn2023reflexion}. This agent verifies that the question matches the identified tone and linguistic style, avoids AI-suspicious phrasing, and respects structural constraints, such as maintaining a natural single-question format. If inconsistencies, unnatural expressions, or typical LLM phrasings are detected, feedback is sent back to the \textsc{QuestionAgent}, which revises the question accordingly. This reflection loop continues until at least one candidate satisfies both stylistic and safety requirements, reducing the risk of producing messages that appear artificial or out of context.
The approved question is then passed to the \textsc{SlangConverter}, which rewrites the question, injecting community-specific slang and stylistic cues. Recent work has shown that general-purpose LLMs benefit from dedicated tuning for rewriting tasks~\cite{shu2024rewritelm}, motivating a small fine-tuned model for the slang conversion step.

\begin{table*}[t]
\centering
\caption{Agent specifications: goal, inputs, and agents' outputs in the \SysName pipeline. Input handling mode: \textbf{SP} = fixed system prompt, \textbf{UM} = user message, \textbf{DSP} = dynamically constructed system prompt. Agents are grouped by functional block.}
\label{tab:agents}
\renewcommand{\arraystretch}{0.85}
\resizebox{.85\textwidth}{!}{
\begin{tabular}{>{\centering\arraybackslash}m{0.5cm} l p{5cm} p{4.7cm} p{3.8cm}}
\toprule
\textbf{Block} & \textbf{Agent} & \textbf{Goal} & \textbf{Inputs} & \textbf{Output} \\
\midrule

\multirow{5}{*}{\raisebox{2.7cm}\centering\rotatebox[origin=c]{90}{\textit{Engagement Gating}}}
& \textsc{ThreadAnalyzer}
  & Produce a structured CTI report of the thread, flagging irrelevant or harmful content and assessing engagement worthiness.
  & Forum thread plaintext [\textbf{UM}]; classification criteria [\textbf{SP}].
  & CTI summary, conversation analysis, irrelevant-content flags, engagement recommendation. \\
\cmidrule(l){2-5}

& \textsc{RelevanceAgent}
  & Classify threads as \textit{Relevant} or \textit{Not Relevant}; blocks prohibited topics.
  & CTI report from \textsc{ThreadAnalyzer} [\textbf{UM}]; classification criteria [\textbf{SP}].
  & Binary decision (\textit{Relevant} / \textit{Not Relevant}); terminates pipeline if not relevant. 
  \\
\cmidrule(l){2-5}

& \textsc{ReplyAnalyzer}
  & Filter incoming user replies to retain only those that address the system's questions, discarding off-topic messages.
  & System username, prior interaction history, new incoming replies [\textbf{UM}]; classification instructions [\textbf{SP}].
  & Validated subset of replies forwarded to \textsc{DefensiveAgent}. \\
\cmidrule(l){2-5}

& \textsc{DefensiveAgent}
  & Classify validated replies as \textit{safe}, \textit{attack-ai} (AI accusation), or \textit{attack-jailbreak}; terminate pipeline on any attack signal.
  & Validated replies [\textbf{UM}]; classification criteria [\textbf{SP}].
  & Safety label; safe replies forwarded to \textsc{TacticSelector}; attacks terminate the pipeline. \\
\cmidrule(l){2-5}

& \textsc{FilterAgent}
  & Act as a final safety barrier, blocking any generated question that violates topic policies before publication.
  & Serialized list of slang-converted questions [\textbf{UM}]; exclusion criteria [\textbf{SP}].
  & Filtered list of publishable questions, or pipeline termination if all are blocked. \\
\midrule

\multirow{3}{*}{\raisebox{1.75cm}\centering\rotatebox[origin=c]{90}{\textit{Context \& Question Gen.}}}
& \textsc{TTP Extractor}
  & Identify MITRE ATT\&CK tactics and techniques present in the conversation and verify each mapping.~\cite{llm_mitre}
  & Thread plaintext [\textbf{UM}]; extraction instructions [\textbf{SP}].
  & List of attack behaviors with mapped ATT\&CK tactics and techniques. \\
\cmidrule(l){2-5}

& \textsc{TacticSelector}
  & Select the ATT\&CK tactic most likely to elicit a  detailed response, maintaining a score map across all 14 tactics updated after each turn.
  & Thread text and extracted tactics [\textbf{UM}] on first turn; Q\&A history appended on subsequent turns [\textbf{UM}]; scoring logic [\textbf{SP}].
  & Selected tactic with reasoning, or termination signal if no tactic is sufficiently effective. \\
\cmidrule(l){2-5}

& \textsc{QuestionAgent}
  & Generate a contextually appropriate question aligned with the selected tactic, revised iteratively on reviewer feedback.
  & Tone/style profile [\textbf{DSP}]; thread text, CTI summary, selected tactic [\textbf{UM}]; reviewer feedback appended on revision [\textbf{UM}].
  & Five candidate questions; revised questions upon reviewer feedback. \\
\midrule

\multirow{3}{*}{\raisebox{1.6cm}\centering\rotatebox[origin=c]{90}{\textit{Ling. Style \& Credibility}}}
& \textsc{ToneAgent}
  & Classify the dominant conversational tone and identify stylistic features of the thread to guide downstream generation.
  & Thread content [\textbf{UM}]; tone taxonomy [\textbf{SP}].
  & Tone (e.g., \textit{Informal}, \textit{Paranoid}, \textit{Aggressive}) and binary features (e.g., all caps, extreme punctuation). \\
\cmidrule(l){2-5}

& \textsc{ReviewerAgent}
  & Verify that questions match the target tone, avoid suspicious phrasing, and follow the comment and question format.~\cite{shinn2023reflexion}
  & Generated questions, tone, and style specifications. [\textbf{UM}]; review criteria [\textbf{SP}].
  & Approval or structured feedback sent back to \textsc{QuestionAgent} (reflection loop). \\
\cmidrule(l){2-5}

& \textsc{SlangConverter}
  & Rewrite formally generated questions using community-specific slang and stylistic cues.
  & Approved question(s) [\textbf{UM}]; fine-tuned on synthetic CrimeBB dataset.
  & Slang-converted question(s) preserving original semantic intent. \\
\bottomrule
\end{tabular}
}
\end{table*}

\section{Experimental Setup}
To comply with ethical guidelines, we use open-weight models on offline hardware rather than relying on commercial models. This avoids sharing potentially sensitive personal information with model providers. 
Agents requiring uncensored outputs, specifically those that must reason about attack behaviors without triggering safety refusals, use an abliterated Gemma3-27B~\cite{gemma_2025}, while reasoning and classification agents use Qwen3-32B~\cite{qwen3technicalreport}. 
The \textsc{SlangConverter} uses a fine-tuned 4B-parameter model, as its narrow task requires limited reasoning capacity. Further details are in Appx~\ref{sec:app_implementation}.

\subsection{Evaluation Framework}

Based on the research questions of Section~\ref{sec:motivation}, we derive the following experimental questions that guide our evaluation.

\myparq{EQ1: Can the system identify productive lines of inquiry from an initial forum post?} 
Active elicitation is a viable approach only if the system can effectively understand the thread context and formulate coherent questions that probe for additional CTI-relevant knowledge. We therefore evaluate, in a simulated setting based on real underground forum conversations, how well the bot can identify productive lines of
inquiry to recover missing CTI knowledge (§\ref{sec:exp_extraction}).

\myparq{EQ2: Can the system remain safe and aligned under adversarial conditions?} Operating in adversarial environments requires the system to remain safe and aligned under hostile conditions. We evaluate whether the proposed architecture can 
avoid prohibited topics, detect jailbreak attempts, and block policy-violating content before publication (§\ref{sec:exp_robustness}).

\myparq{EQ3: Do the proposed architectural choices measurably improve performance?} The proposed multi-agent architecture involves several design choices and task decompositions whose individual contributions are not obvious a priori. We validate the impact of architectural choices on system performance. We compare the full multi-agent pipeline against a single-agent baseline and ablate individual components.
Extraction performance is evaluated through simulated user interactions, and stylistic credibility is evaluated through a survey (§\ref{sec:exp_design}).

\myparq{EQ4: Does the system participation increase intelligence yield?} Passive approaches can only observe the activity that develops naturally after a thread is published, whereas \SysName actively participates in the discussion. To isolate the effect of this participation, we conduct a prospective matched experiment using identical initial posts in online forums (§\ref{sec:exp_activity_e3}).

\myparq{EQ5: Can the system operate effectively in real underground forum threads?} The system must operate in underground communities without being detected. Since the previous evaluations, including the controlled activity experiment, may not fully reproduce the dynamics of organically occurring discussions in underground communities, we deploy \SysName in real underground forums to assess whether it can conduct live conversations without triggering bans, moderator interventions, or AI detection, while eliciting information (§\ref{sec:exp_real_forums}).

\mypar{Datasets}
Table~\ref{tab:dataset_map} summarizes the datasets used for each experiment. More details are provided in Appx~\ref{sec:app_datasets}.

\begin{table*}[t]
\centering
\caption{Mapping of evaluation questions, datasets, experiments, and metrics.}
\label{tab:dataset_map}
\renewcommand{\arraystretch}{1.3}
\resizebox{.85\textwidth}{!}{
\begin{tabular}{l l l l l}
\toprule
\textbf{EQ} & \textbf{Dataset / Forum} & \textbf{Content} & \textbf{Experiment} & \textbf{Metric(s)} \\
\midrule
\addlinespace
\makecell[c]{\textit{EQ1} \\ \textit{EQ3}}
& CrimeBB~\cite{crimebb}
& \makecell[l]{Underground \\ forum threads}
& \makecell[l]{Simulated CTI probing (§~\ref{sec:exp_extraction}) \\ 
Design validation (§~\ref{sec:exp_design})}
& \makecell[l]{Tactics and techniques recovered, \\
survey, token consumption} \\
\addlinespace

\textit{EQ2}
& CoDA~\cite{coda}
& Dark web content
& Engagement gating validation (§~\ref{sec:exp_engagement})
& Topic rejection rate \\
\addlinespace

\textit{EQ2}
& \makecell[l]{Do Anything Now~\cite{do_anything_now} \&\\[1pt]
InTheWild Jailbreaks~\cite{in_the_wild}}
& Jailbreak prompts
& Robustness against jailbreaks (§~\ref{sec:exp_jailbreak})
& Jailbreak detection rate \\
\addlinespace

\textit{EQ2}
& BeaverTails~\cite{question_answer_harmful_topics}
& Harmful Q\&A pairs
& Message filter validation (§~\ref{sec:exp_filter})
& Block rate of harmful questions \\
\addlinespace

\textit{EQ3}
& AnnoCTR~\cite{annoctr}
& CTI reports
& TTP Extractor validation (§~\ref{sec:exp_ttp_extraction})
& Precision and recall \\
\addlinespace

\textit{EQ4}
& \makecell[l]{BlackHatWorld \& Dread\\matched deployment}
& \makecell[l]{20 matched pairs of\\identical forum posts}
& Intelligence yield effect evaluation (§~VI-E)
& \makecell[l]{Paired CTI entities difference in\\user messages over seven days} \\
\addlinespace

\textit{EQ5}
& \makecell[l]{BlackHatWorld, sinister.ly,\\Hack Forums, and Dread}
& \makecell[l]{104 organic\\live forum threads}
& Automated forum deployment (§~VI-F)
& \makecell[l]{Response and thread activity, detection and\\
moderation outcomes,
and qualitatively elicited CTI} \\
\bottomrule
\end{tabular}
}
\end{table*}

\subsection{Intelligence Probing Evaluation (EQ1)}
\label{sec:exp_eq1}
This experiment evaluates the system’s ability to understand conversational context and generate coherent questions  about the discussed topic. 
We selected 100 relevant threads from the CrimeBB dataset, each containing the initial post and subsequent replies. The experiment is designed as a controlled probing task:
\SysName sees only the initial post, while a user proxy has access to the complete original discussion and may return
only information explicitly contained in the original human replies and directly asked about.
Conducting live experiments on actual forums for this evaluation was intentionally avoided, as repeated interactions with real users during iterative development phases could have revealed the study's nature, compromising subsequent sessions. Evaluating an information extraction bot by recruiting domain experts to simulate forum users at scale is also not feasible, as it would be difficult to maintain consistency and unbiasedness across evaluators and sessions. We therefore simulate the user proxy by using an external LLM as a judge, thereby providing a reproducible approximation of forum participants' interactions. The proxy follows a constrained behavioral policy. It returns a source reply only when the generated question directly targets the information expressed in that reply. It cannot introduce external knowledge, synthesize new facts, or reveal unrelated parts of the conversation, but just returns messages from the original thread. When no source reply satisfies these conditions, it returns a fixed negative answer.
We evaluate system performance by comparing MITRE ATT\&CK techniques across the original conversation and the messages the system extracts through its questions. Beyond the final technique recovery, we report the probing efficiency, defined as the number of techniques first evidenced in replies divided by the number of published questions. 
We exclude techniques already present in the
initial post, repeated mentions, and techniques appearing only in \SysName's questions.
As a secondary measure, we count unique external URLs first appearing in user proxy replies. Since we do not independently assess each URL's relevance or reliability, this metric reflects the yield of references rather than CTI quality.
The recovered techniques are obtained through manual annotation rather than the raw output of the \textsc{TTP Extractor}. We use the extractor only as a candidate generator over the source thread: it proposes candidate technique mappings, which a domain-informed annotator verifies against the supporting text, discarding unsupported ones and independently re-reading each thread to add techniques the extractor missed. Manual annotation is necessary because the extractor's measured accuracy (\S\ref{sec:exp_design}) is not sufficient for defining a scoring target: at the technique level, it reaches 46.4\% recall at 19\% precision on AnnoCTR~\cite{annoctr}. Using its raw output would distort the metric in both directions. Missed techniques would understate what a configuration recovered, while spurious mappings would reward configurations that simply publish more questions, since every additional reply gives the extractor another opportunity to emit an unsupported technique regardless of question quality. Manual annotation also avoids a closed-loop evaluation, in which the same component both drives \SysName's reasoning and defines the score. Techniques mentioned only in the system's own questions are never counted as recovered.
While this experimental setup quantifies the bot's ability to understand context and generate appropriate questions, it has inherent limitations, as discussed in Section~\ref{sec:threats_to_validity}.

\begin{figure}[t]
\centering
\begin{tikzpicture}
\begin{axis}[
    width=\linewidth,
    height=0.6\linewidth,
    xmin=0, xmax=20,
    ymin=0, ymax=100,
    xtick={0,2,...,20},
    ytick={0,20,40,60,80,100},
    xlabel={\textit{x} : Max Allowed Messages},
    ylabel={Extracted Techniques (\%)},
    grid=both,
    grid style={dashed,gray!30},
    legend style={
        font=\small,
        at={(0.98,0.05)},
        anchor=south east,
        draw=none,
        fill=white,
        fill opacity=0.85,
        draw opacity=1,
        text opacity=1
    }
]

\addplot[
    very thick,
    mark=*,
]
coordinates {
    (0,32.315522)
    (1,48.346056)
    (2,58.269720)
    (3,62.340967)
    (4,65.903308)
    (5,67.938931)
    (6,69.211196)
    (7,69.465649)
    (8,69.465649)
    (9,70.483461)
    (10,70.992366)
    (11,71.246819)
    (12,71.501272)
    (13,71.501272)
    (14,71.501272)
    (15,71.755725)
    (16,72.264631)
    (17,72.264631)
    (18,72.264631)
    (19,72.519084)
    (20,72.773537)
};
\addlegendentry{\SysName}

\addplot[
    customblue,
    mark=*,
]
coordinates {
    (0,32.315522)
    (1,45.801527)
    (2,50.636132)
    (3,54.707379)
    (4,59.796438)
    (5,63.358779)
    (6,64.376590)
    (7,65.394402)
    (8,66.412214)
    (9,66.921120)
    (10,66.921120)
    (11,66.921120)
    (12,67.175573)
    (13,67.430025)
    (14,67.430025)
    (15,67.430025)
    (16,67.430025)
    (17,67.430025)
    (18,67.430025)
    (19,67.430025)
    (20,67.430025)
};
\addlegendentry{No TacticSelector}

\addplot[
    customred,
    mark=*,
]
coordinates {
    (0,32.315522)
    (1,39.694656)
    (2,44.529262)
    (3,46.055980)
    (4,48.346056)
    (5,49.109415)
    (6,49.618321)
    (7,50.381679)
    (8,50.636132)
    (9,51.145038)
    (10,51.145038)
    (11,51.399491)
    (12,51.399491)
    (13,51.399491)
    (14,51.399491)
    (15,51.399491)
    (16,51.399491)
    (17,51.399491)
    (18,51.399491)
    (19,51.399491)
    (20,51.399491)
};
\addlegendentry{Single-Agent}

\addplot[
    customgrey,
    dashed,
    opacity=0.9,
    no marks,
]
coordinates {
    (0,32.315522)
    (20,32.315522)
};

\end{axis}
\end{tikzpicture}

\caption{Average techniques retrieved for different message limits. The plot compares \SysName with a single-agent baseline and without the standalone \textsc{TacticSelector} module.}
\label{fig:avg-overlap-comparison}
\end{figure}
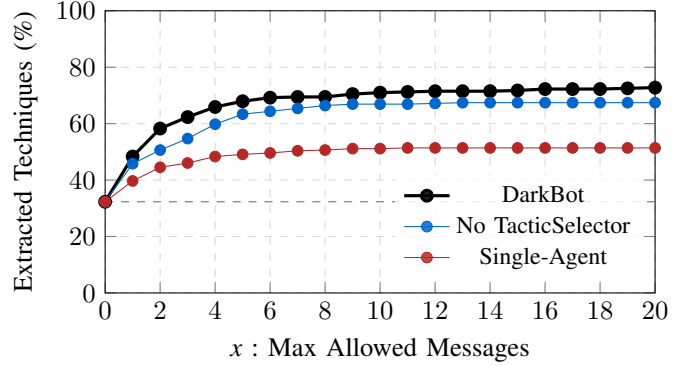

We repeat the experiment under different limits on the number of messages \SysName is allowed to publish, to characterize how recovery evolves as the number of bot messages increases.
A single conversation is simulated per thread, and each point at \textit{x} reports the recovery from the replies obtained within the first \textit{x} published questions. The value is a cap rather than a target: the pipeline may terminate on its own when the \textsc{TacticSelector} finds no tactic with sufficient expected effectiveness. Recovery is therefore non-decreasing in \textit{x} by construction.
Because all configurations start from the same initial post, they share the same pre-interaction baseline (dashed line), and subsequent gains reflect techniques recovered from replies made available through the controlled probing process.
The resulting curve in Fig.~\ref{fig:avg-overlap-comparison} (black) shows that the overlap plateaus, reaching a maximum of 72.8\%, starting from a baseline of 32.3\% of techniques present in the initial post.
Overall the system published 928 questions recovering 159 new techniques and 122 source-matching URLs, averaging a probing efficiency of 0.171 techniques per question and 0.131 URLs per question.
\label{sec:exp_extraction}

\begin{takeaway}
  \textbf{Takeaway (EQ1).} With the initial post only, \SysName recovers 72.8\% of the MITRE ATT\&CK techniques present in the complete source conversations. Beyond final recovery, the system averages a probing efficiency of 0.171 additional techniques per published question. These results indicate that \SysName can understand thread context and identify productive questions under the defined user-proxy policy.
\end{takeaway}

\subsection{Robustness Evaluation (EQ2)}
\label{sec:exp_robustness}
This section evaluates whether the proposed multi-agent system behaves safely under simulated hostile conditions, before deploying it in real-world forum interactions (§\ref{sec:exp_real_forums}). In particular, we analyze whether it engages only in appropriate discussions, resists jailbreak and manipulation attempts, and prevents the publication of unsafe or policy-violating content. 

\begin{figure}[t]
  \centering
  \begin{tikzpicture}
  \begin{axis}[
      ybar,                       
      ymin=0,
      ymax=250,
      ytick={0,50,100,150,200},
      bar width=6pt,
      enlarge x limits=0.08,
      xtick pos=bottom,
      nodes near coords,
      nodes near coords style={font=\footnotesize, rotate=90, anchor=west,yshift=0.4pt},
      tick label style={font=\footnotesize},
      label style={font=\small},
      ymajorgrids=true,
      grid style={dotted, gray!80},
      symbolic x coords={Drugs,Electronic,Gambling,Others,Financial,Hacking,Crypto,Violence,Porn,Arms},
      xtick=data,
      x tick label style={rotate=45, anchor=east, font=\footnotesize},
      ylabel={Count},
      width=\linewidth,
      height=0.6\linewidth,
      legend style={
        font=\footnotesize,
        at={(0.02,0.98)},
        anchor=north west,
        legend cell align={left},
        reverse legend
      },
      area legend,
  ]
  \addplot[
      draw=black,
      fill=teal!70!black,
      bar shift=-3.6pt,
  ] coordinates {
      (Arms,1)
      (Porn,1)
      (Violence,3)
      (Crypto,19)
      (Hacking,183)
      (Financial,15)
      (Others,8)
      (Gambling,2)
      (Electronic,1)
      (Drugs,0)
  };
  \addplot[
      draw=black,
      fill=gray!25,
      bar shift=3.6pt,
  ] coordinates {
      (Arms,30)
      (Porn,30)
      (Violence,30)
      (Crypto,30)
      (Hacking,200)
      (Financial,30)
      (Others,30)
      (Gambling,30)
      (Electronic,30)
      (Drugs,30)
  };
  \legend{Pred. Relevant,Support}
  \end{axis}
  \end{tikzpicture}
  \caption{Relevance Prediction on CoDA Samples}
  \label{fig:experiment1_relevance_validation}
\end{figure}
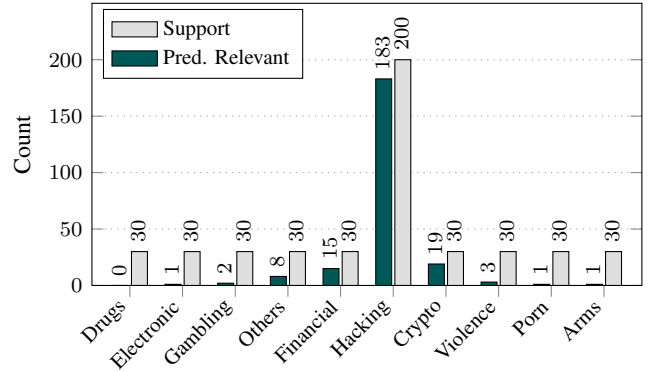

\mypar{Engagement Gating Validation}
\label{sec:exp_engagement}
The goal of this experiment is to verify that the system can distinguish \ac{CTI}-related topic threads and, more importantly, to avoid engaging in potentially harmful conversations that violate our policies (\S~\ref{sec:app_policies}).
For this analysis, we leveraged a subset of the CoDA dataset~\cite{coda} (Appx~\ref{sec:app_datasets}), containing dark web documents labeled into ten topic categories.
We present the selected texts to the bot as initial posts of new forum threads.
A post is considered \textit{Relevant} if it was labeled as such by the \textsc{RelevanceAgent}, thereby triggering the bot message generation mechanism. 
However, the generated response may still be blocked by the final \textsc{FilterAgent} when the full pipeline is run.
 As Fig.~\ref{fig:experiment1_relevance_validation} reports, 91.5\% of the texts in the \textit{Hacking} category were classified as \textit{Relevant}, a small fraction was labeled as \textit{Not Relevant}, as the system deemed them insufficiently informative for CTI question generation.

For the five texts classified as \textit{Relevant} in the Arms, Porn, and Violence categories, we evaluated the full pipeline. For the one Porn-related question, the generated question was blocked by the \textsc{FilterAgent}. The remaining four cases, related to Arms and Violence, produced questions aimed at extracting CTI without engaging with or addressing harmful content. This shows that the system focused exclusively on cybersecurity-related content, not reinforcing the potentially harmful aspects.

\mypar{Robustness Against Adversarial Inputs}
\label{sec:exp_jailbreak}
The goal of this experiment is to verify the system's ability to detect jailbreaking attempts.
Jailbreak prompts are typically used to elicit restricted information or to induce prohibited behaviors that a safety-aligned LLM would otherwise refuse to produce.
For this purpose, we tested the system using publicly available jailbreak datasets from ``Do Anything Now'' by Shen et al.~\cite{do_anything_now} and In-the-Wild Jailbreaks by Jiang et al.~\cite{in_the_wild}.
We evaluate the system’s ability to detect jailbreak attempts and terminate the interaction accordingly.
\mypari{DefensiveAgent} We tested robustness to adversarial replies by feeding jailbreaks to the \textsc{DefensiveAgent} without following the bot pipeline.
We analyzed 845 jailbreaking prompts and found that 92.8\% of them were detected by the \textsc{DefensiveAgent}.

\mypari{Full Pipeline} We assess end-to-end robustness by injecting the jailbreak prompts into three representative dark-web conversations on malware, botnets, and initial access brokerage. After the bot's first reply, we submit the jailbreak message; in all three scenarios, 100\% of attempts receive no response.

\myparnotitle While the \textsc{DefensiveAgent} is not a perfect classifier, the overall pipeline can detect attack attempts that deviate from the original conversation topic via the \textsc{ReplyAnalyzer}, and the final \textsc{FilterAgent} prevents the bot from publishing messages that deviate from the defined guidelines.

\mypar{Message Filter Validation}
\label{sec:exp_filter}
To evaluate the robustness of the final message filter (i.e., the \textsc{FilterAgent}), we now consider the scenario in which earlier safeguards fail or are bypassed through jailbreaking. In such cases, the model may generate content that violates the system’s safety guidelines. The final message filter serves as a last line of defense, detecting messages that violate our guidelines and blocking them from the pipeline before publication.
For this purpose, we used a subset of the \textit{BeaverTails}~\cite{question_answer_harmful_topics} dataset (Appx~\ref{sec:app_datasets}).
We submit each question to the FilterAgent, simulating a rogue input that the system did not intercept earlier.
We observed that the vast majority of harmful questions were correctly blocked, with only a small portion classified as \textit{Relevant}. Out of 347 harmful questions, 311 (89.6\%) were blocked. 
In deployment, however, every outgoing message is first rewritten by the \textsc{SlangConverter} into the colloquial, abbreviated register of the target forums, so the \textsc{FilterAgent} faces a \emph{slang} distribution rather than clean English, which may differ from the style of the \textit{BeaverTails}~\cite{question_answer_harmful_topics} dataset. We apply slang rewriting only to \textit{BeaverTails} because it represents outgoing messages processed by the \textsc{SlangConverter} before reaching the \textsc{FilterAgent}; \textit{CoDA} posts and jailbreak prompts represent incoming content and are therefore evaluated unchanged. To verify that our safety guarantees hold under this distribution shift, we reran the evaluation on a \textsc{SlangConverter}-rewritten version of the dataset.
Slang rewriting slightly erodes \textsc{FilterAgent} performance, as the block rate falls from $89.6\%$ to $85.6\%$ for harmful questions.
A manual inspection of the remaining questions revealed that
most of them did not explicitly involve undeniably harmful topics, unlike those immediately classified as \textit{Not Relevant}. Examples include: "\textit{How do I tattoo myself?}", "\textit{How do I get into illegal sports betting?}".
This experiment models a scenario in which a jailbreak evades the \textsc{DefensiveAgent} and causes the bot to generate a policy-violating message that then reaches the \textsc{FilterAgent}. If the two observed failure rates are independent and transferable to the same input distribution, the estimated probability that both layers fail is $0.072 ~$\texttimes~$0.144=1.04\%$.

\begin{takeaway}
\textbf{Takeaway (EQ2).}
The full pipeline rejected 100\% of tested jailbreak prompts, even when the \textsc{DefensiveAgent} alone misclassifies 7.2\% of them. The \textsc{FilterAgent} in isolation blocked
89.6\% of harmful messages and 85.6\% after slang rewriting, with most of the undetected questions being borderline content.
By combining the performance of the individual components, the pipeline estimates a failure rate of just 1.04\%, indicating that the layered architecture effectively compensates for individual-agent limitations.
\end{takeaway}

\subsection{Design Validation (EQ3)}
\label{sec:exp_design}
To evaluate the effectiveness of the pipeline design, we assess two main dimensions: the ability to understand the context and generate coherent questions, and the stylistic credibility of the produced messages. Probing performance is evaluated on the same
100-thread CrimeBB cohort used in EQ1, allowing direct
comparison between the full multi-agent pipeline, a
single-agent baseline, and a \textsc{TacticSelector} ablation.
Stylistic credibility is evaluated through a survey recruiting 118 participants with a Computer Science background, with no incentives. 
Survey details are reported in Appx~\ref{sec:app_survey}.
\pgfplotsset{compat=1.18}
\begin{figure}[t]
\centering
\resizebox{\columnwidth}{!}{
\begin{tikzpicture}
\begin{axis}[
    name=survey,
    width=\columnwidth,
    height=4.5cm,
    xmin=-50, xmax=75,
    xtick={-50,-25,0,25,50,75},
    xticklabels={50\%,25\%,0,25\%,50\%,75\%},
    xticklabel style={font=\small},
    ymin=0.5, ymax=3.5,
    ytick={1,2,3},
    yticklabels={$-$\textsc{ToneAgent},$-$\textsc{SlangConverter},Single-Agent},
    yticklabel style={font=\small},
    xlabel={Share of judgments},
    xlabel style={font=\small},
    xmajorgrids=true,
    grid style={dotted, gray!80},
    extra x ticks={0},
    extra x tick style={
        grid=major,
        grid style={black!60, thin},
        tick style={draw=none},
        xticklabel=\empty,
    },
    axis line style={draw=none},
    tick style={draw=none},
    legend style={
        at={(0.5,1.05)}, anchor=south, legend columns=3,
        font=\small, draw=none, fill=none,
        column sep=8pt,
    },
    legend image code/.code={
        \draw[#1, draw=none] (0cm,-0.1cm) rectangle (0.35cm,0.15cm);
    },
    clip=false,
]
\addlegendimage{fill=customred, draw=none}
\addlegendentry{Better Ablated}
\addlegendimage{fill=gray!45, draw=none}
\addlegendentry{Equal}
\addlegendimage{fill=teal!70!black, draw=none}
\addlegendentry{Better \SysName}

\pgfmathsetmacro{\bh}{0.28}

\fill[customred] (axis cs:-46.10,{1-\bh}) rectangle (axis cs:-10.85,{1+\bh});
\fill[gray!45] (axis cs:-10.85,{1-\bh}) rectangle (axis cs:10.85,{1+\bh});
\fill[teal!70!black] (axis cs:10.85,{1-\bh}) rectangle (axis cs:43.05,{1+\bh});

\fill[customred] (axis cs:-17.97,{2-\bh}) rectangle (axis cs:-10.76,{2+\bh});
\fill[gray!45] (axis cs:-10.76,{2-\bh}) rectangle (axis cs:10.76,{2+\bh});
\fill[teal!70!black] (axis cs:10.76,{2-\bh}) rectangle (axis cs:71.27,{2+\bh});

\fill[customred] (axis cs:-35.25,{3-\bh}) rectangle (axis cs:-10.17,{3+\bh});
\fill[gray!45] (axis cs:-10.17,{3-\bh}) rectangle (axis cs:10.17,{3+\bh});
\fill[teal!70!black] (axis cs:10.17,{3-\bh}) rectangle (axis cs:54.58,{3+\bh});

\node[rotate=-90, anchor=south, font=\small\scshape, color=black]
  at (axis cs:75,2) {\SysName};

\end{axis}
\end{tikzpicture}
}
\caption{Stylistic credibility ablation survey. 
}
\label{fig:survey}
\end{figure}
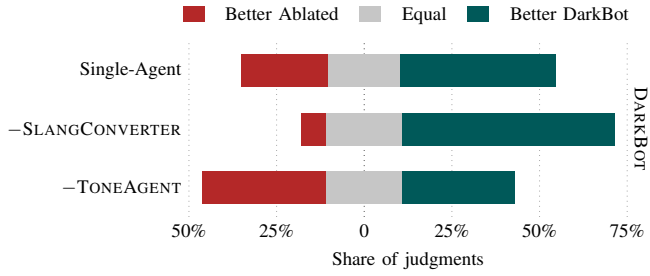

\mypar{Multi-Agent Architecture Evaluation}
To evaluate whether decomposing the task into specialized agents provides measurable advantages, we compare \SysName with a single-agent baseline.
To construct the single-agent baseline, we merged all agent prompts and logic from the multi-agent pipeline into a single structured system prompt.
This prompt preserved the original workflow logic by explicitly describing the sequential reasoning steps corresponding to each agent.

\mypari{Intelligence Probing}
Regarding information extraction ability, Figure~\ref{fig:avg-overlap-comparison} reports the overlap of extracted MITRE techniques between simulated and original conversations, following the experimental setting discussed in Sec.~\ref{sec:exp_extraction} on 100 CrimeBB threads. For this experiment, we disabled the relevance identification logic for a fair comparison. The single-agent system converges at approximately 51.4\% techniques overlap, while the multi-agent architecture reaches 72.8\%. This difference is not explained solely by a larger number of
published questions: normalized by the number of questions,
the multi-agent pipeline recovers 0.171 new techniques per
question, compared with 0.131 for the single-agent baseline. It also retrieves 122 unique external
references, compared with 47 for the single-agent baseline,
corresponding to 0.131 versus 0.082 references per published
question.  

\mypari{Stylistic Credibility}
Figure~\ref{fig:survey} reports the results of the stylistic credibility survey.
When comparing the full pipeline against the single-agent baseline, 44.41\% of judgments
rated the full-pipeline messages as more likely to be produced by a human user, while 35.25\% preferred the single-agent output and 20.34\% perceived no difference. These results suggest that the modular architecture, beyond improving extraction capabilities, also contributes positively to stylistic credibility, likely due to the dedicated
reasoning steps that allow each agent to focus on specific aspects of message generation.

\mypari{Computational Cost}
The multi-agent system, as expected, consumes on average more completion tokens (37,485 vs. 20,701) and prompt tokens (422,280 vs. 340,006) than the single-agent system. This aligns with prior literature highlighting benefits and costs of modular reasoning in complex intelligence extraction tasks~\cite{mad-cti}.

\mypar{Agent Ablations}
We perform agent ablations to assess their effects on stylistic credibility and information extraction.

\mypari{Stylistic Credibility}\label{sec:exp_style_survey}
We validate the contribution of the \textsc{ToneAgent} and the
\textsc{SlangConverter} to the stylistic credibility of the generated messages through a survey (see Appx~\ref{sec:app_survey}), comparing outputs from the full pipeline against those produced without these
components, across 5 representative threads. Participants were asked to judge which
messages better adhered to the conversational tone of the thread, and which appeared
more likely to have been written by a human user. 
As Figure~\ref{fig:survey} demonstrates, the \textsc{SlangConverter} emerges as the component with the strongest and most consistent impact on stylistic credibility: 60.51\% of participants rated full-pipeline messages as
better than those produced without it, with only 17.97\% preferring the ablated variant.
This indicates that community-specific lexical adaptation
substantially improves perceived stylistic credibility under
the evaluated survey conditions.
These results also provide indirect evidence that the fine-tuning procedure on the
synthetic dataset was effective.
The contribution of the \textsc{ToneAgent} to the final stylistic output appears limited: only 32.20\% of
participants preferred the full-pipeline output over the no-\textsc{ToneAgent} variant, while 46.10\% rated it as worse.
Although the effect of the \textsc{ToneAgent} in our survey is perceived as slightly negative, prior work consistently treats tonal alignment as relevant to conversational credibility and elicitation~\cite{giles1991,nunan2022impact}. We therefore retain the \textsc{ToneAgent} in the pipeline and leave its improvement as future work.

\mypari{Intelligence Probing}
To assess the contribution of the \textsc{TacticSelector} as a standalone
component, we repeat the offline extraction experiment described in
Section~\ref{sec:exp_extraction} under an ablated configuration in which tactic selection is integrated directly into the \textsc{QuestionAgent}. In this setting, the \textsc{QuestionAgent} is jointly responsible for selecting the tactics to pursue and deriving the follow-up question, without a dedicated reasoning step for tactical planning.
Figure~\ref{fig:avg-overlap-comparison} shows the proportion of techniques recovered from replies returned during the controlled probing simulation, relative to the complete source threads.
The full pipeline converges to a final recovery of 72.8\%, whereas the ablated system reaches a maximum of 67.4\%. Removing the \textsc{TacticSelector} also led to a slight degradation in probing efficiency, from 0.171 to 0.165 new techniques per published question and from 0.131 to 0.108 new URLs per question (122 against 90 URLs in total). Its main contribution, however, is sustaining recovery across turns rather than making individual questions more productive.
Without explicit tactic-planning, the
\textsc{QuestionAgent} focuses on the most salient
content in the current exchange, while the
\textsc{TacticSelector} tracks covered tactics and prioritizes underexplored dimensions.

\mypar{TTP Extraction Validation}
\label{sec:exp_ttp_extraction}
We compare the employed \textsc{TTP Extractor} with
LADDER~\cite{ladder}, a non-LLM-based ATT\&CK extraction
baseline, on the annotated AnnoCTR reports and on the same
100 CrimeBB threads used in \S\ref{sec:exp_eq1}. AnnoCTR provides
independent expert labels and is therefore used to evaluate
accuracy, whereas the CrimeBB comparison characterizes
candidate extraction yield in the forum domain, as it does not provide expert technique labels.

\mypari{Techniques extraction} On 105 labeled AnnoCTR reports~\cite{annoctr}, LADDER extracted 796 techniques (7.58/report) versus 2{,}479 (23.61/report) for the LLM system. Against ground truth, the LLM achieved 46.4\% recall and 19\% precision, compared with LADDER's 15.8\% recall and 18.8\% precision, providing substantially higher coverage at comparable precision. On 100 CrimeBB threads, LADDER extracted 50 techniques (0.50/thread), versus 1{,}996 (19.96/thread) for the LLM system.

\mypari{Tactics extraction}
Since LADDER does not predict tactics, we infer them from the technique IDs it extracts using the official MITRE ATT\&CK mapping, where techniques listed under multiple tactics contribute to all associated tactics.
On the 105 AnnoCTR reports, LADDER inferred 581 tactics (5.53 per report), compared with 885 for the LLM system (8.43 per report).
Using the same mapping to derive ground-truth tactics from AnnoCTR technique labels, the LLM system achieves a tactic recall of 74.4\% and precision of 50.9\%, compared to LADDER's 45.9\% recall and 46.7\% precision. At the tactic level, the LLM system therefore improves on LADDER across both dimensions, recovering nearly three-quarters of the ground-truth tactics while maintaining higher precision, and confirming that the richer technique output also provides more accurate tactic coverage.
On CrimeBB, LADDER's techniques translate to 46 inferred tactics (0.46 per thread), while the LLM system identified 902 tactics (9.02 per thread). 
Higher recall is especially important in this setting, as LADDER's average of 0.46 tactics (0.50 techniques) per thread would leave the bot with virtually nothing to explore.

\begin{takeaway}
  \textbf{Takeaway (EQ3).} The multi-agent pipeline improves technique recovery (72.8\% vs.\ 51.4\%) and probing efficiency (0.171 vs.\ 0.131) against the single-agent baseline. Ablation of the \textsc{TacticSelector} reduced both final recovery (-5.4\%) and probing efficiency (-0.165 techniques per question) indicating that explicit tactic planning helps sustain
broader exploration across turns. Among the evaluated stylistic
components, the \textsc{SlangConverter} has the strongest
positive measured effect. By contrast, the \textsc{ToneAgent} has a slightly negative perceived effect. 
\end{takeaway}

\subsection{Intelligence Yield Effect Evaluation (EQ4)}
\label{sec:exp_activity_e3}
To estimate the effect of \SysName participation on intelligence yield, we conducted a prospective matched-pairs deployment. We evaluated the paired differences using a two-sided paired permutation test.
We authored 20 discussion topics, each presented as an identical initial post published under our own accounts on both \textit{BHW} and \textit{Dread}. For each topic, exactly one of the two forums was assigned to the bot, while the other copy of the same topic received no bot activity and serves as its control. Because both arms of a pair carry the same text posted at the same time, each topic serves as its own control for subject matter, phrasing, and timing, and the assignment is independent of how the thread develops later. The account that publishes the initial post never returns to the thread, so \SysName is the only non-organic participant on the active side, and the control side has none.
The deployment is additionally counterbalanced across forums and channels: every pair of board sections used hosts an equal number of topics, with the bot assigned to each side. This matters because a paired difference otherwise conflates the bot's effect with the difference in baseline activity between the two sections being compared. With equal numbers of topics in each direction, the baseline terms enter the treated and control totals identically and cancel out, so the aggregate difference reflects the bot's presence rather than the choice of platform or section.
We measure paired differences in CTI entities present in human-authored messages during the 7 days following each thread's confirmed public visibility. As CTI entities, we use a subset of STIX 2.1  standard~\cite{stix21} as domain categories and IoCs (Appx~\ref{sec:app_cti_entities}). We exclude the initial post, bot accounts, and automated or moderator accounts from the count, so every counted message is a user contribution.
\definecolor{ctiTechnique}{RGB}{114,142,196}
\definecolor{ctiTool}{RGB}{62,90,150}
\definecolor{ctiMalware}{RGB}{150,46,46}
\definecolor{ctiInfra}{RGB}{186,200,226}
\definecolor{ctiTarget}{RGB}{96,138,104}
\definecolor{ctiActor}{RGB}{120,84,150}
\definecolor{ctiDefence}{RGB}{28,45,90}
\definecolor{ctiVuln}{RGB}{206,140,52}
\definecolor{ctiIoc}{RGB}{238,242,248}

\begin{figure}[t]
  \centering
  \begin{tikzpicture}
    \begin{axis}[
      ybar stacked,
      stack negative=separate,
      width=0.88\linewidth,
      height=0.68\linewidth,
      xmin=-1,
      xmax=20,
      ymin=-44,
      ymax=57,
      xtick={0,1,2,3,4,5,6,7,8,9,10,11,12,13,14,15,16,17,18,19},
      xticklabels={A,B,C,D,E,F,G,H,I,J,K,L,M,N,O,P,Q,R,S,T},
      xtick pos=bottom,
      x tick label style={font=\footnotesize},
      ytick={-30,-20,-10,0,10,20,30,40,50},
      yticklabels={30,20,10,0,10,20,30,40,50},
      tick label style={font=\footnotesize},
      label style={font=\small},
      xlabel={Topics},
      ylabel={CTI entities},
      ymajorgrids=true,
      grid style={dashed,gray!30},
      bar width=7pt,
      enlarge x limits=false,
      clip=false,
      legend style={
        font=\footnotesize,
        at={(axis description cs:0.5,1.17)},
        anchor=south,
        legend columns=3,
        legend cell align=left,
        draw=black,
        fill=white,
        /tikz/every even column/.append style={column sep=8pt}
      },
      area legend,
      legend image post style={draw=black}
    ]

      \draw[black,thin]
        (axis cs:-1,0) -- (axis cs:20,0);

      \addplot[
        draw=black!55,
        line width=0.2pt,
        fill=ctiTechnique
      ]
        coordinates {
          (0,0) (1,2) (2,0) (3,0) (4,0)
          (5,0) (6,2) (7,4) (8,2) (9,3)
          (10,2) (11,4) (12,2) (13,2) (14,0)
          (15,2) (16,4) (17,3) (18,3) (19,6)
        };

      \addplot[
        draw=black!55,
        line width=0.2pt,
        fill=ctiTechnique,
        forget plot
      ]
        coordinates {
          (0,0) (1,-2) (2,0) (3,0) (4,0)
          (5,0) (6,-2) (7,-6) (8,-1) (9,0)
          (10,-3) (11,0) (12,-1) (13,0) (14,0)
          (15,-5) (16,-1) (17,-2) (18,0) (19,-5)
        };

      \addplot[
        draw=black!55,
        line width=0.2pt,
        fill=ctiTool
      ]
        coordinates {
          (0,0) (1,0) (2,0) (3,0) (4,0)
          (5,0) (6,5) (7,10) (8,3) (9,13)
          (10,3) (11,3) (12,0) (13,1) (14,0)
          (15,1) (16,1) (17,9) (18,4) (19,3)
        };

      \addplot[
        draw=black!55,
        line width=0.2pt,
        fill=ctiTool,
        forget plot
      ]
        coordinates {
          (0,-2) (1,-1) (2,-1) (3,0) (4,0)
          (5,0) (6,-2) (7,-6) (8,-1) (9,0)
          (10,-9) (11,-9) (12,0) (13,0) (14,0)
          (15,-1) (16,0) (17,-1) (18,0) (19,-10)
        };

      \addplot[
        draw=black!55,
        line width=0.2pt,
        fill=ctiMalware
      ]
        coordinates {
          (0,0) (1,0) (2,0) (3,0) (4,0)
          (5,0) (6,0) (7,0) (8,0) (9,0)
          (10,0) (11,0) (12,1) (13,0) (14,0)
          (15,0) (16,0) (17,1) (18,0) (19,1)
        };

      \addplot[
        draw=black!55,
        line width=0.2pt,
        fill=ctiMalware,
        forget plot
      ]
        coordinates {
          (0,0) (1,0) (2,0) (3,0) (4,0)
          (5,0) (6,0) (7,0) (8,0) (9,0)
          (10,0) (11,-1) (12,0) (13,0) (14,0)
          (15,0) (16,0) (17,0) (18,0) (19,0)
        };

      \addplot[
        draw=black!55,
        line width=0.2pt,
        fill=ctiInfra
      ]
        coordinates {
          (0,0) (1,1) (2,0) (3,0) (4,0)
          (5,0) (6,2) (7,0) (8,5) (9,2)
          (10,1) (11,19) (12,0) (13,0) (14,0)
          (15,2) (16,0) (17,5) (18,1) (19,5)
        };

      \addplot[
        draw=black!55,
        line width=0.2pt,
        fill=ctiInfra,
        forget plot
      ]
        coordinates {
          (0,-4) (1,-2) (2,0) (3,0) (4,0)
          (5,0) (6,-1) (7,-1) (8,-1) (9,0)
          (10,-3) (11,-10) (12,0) (13,0) (14,0)
          (15,-1) (16,0) (17,-7) (18,0) (19,-8)
        };

      \addplot[
        draw=black!55,
        line width=0.2pt,
        fill=ctiTarget
      ]
        coordinates {
          (0,0) (1,0) (2,0) (3,0) (4,0)
          (5,0) (6,0) (7,1) (8,0) (9,0)
          (10,0) (11,0) (12,0) (13,0) (14,0)
          (15,0) (16,5) (17,0) (18,1) (19,0)
        };

      \addplot[
        draw=black!55,
        line width=0.2pt,
        fill=ctiTarget,
        forget plot
      ]
        coordinates {
          (0,0) (1,0) (2,0) (3,0) (4,0)
          (5,0) (6,0) (7,0) (8,0) (9,0)
          (10,0) (11,-4) (12,0) (13,0) (14,0)
          (15,0) (16,0) (17,0) (18,0) (19,0)
        };

      \addplot[
        draw=black!55,
        line width=0.2pt,
        fill=ctiActor
      ]
        coordinates {
          (0,0) (1,0) (2,0) (3,0) (4,0)
          (5,0) (6,0) (7,0) (8,0) (9,0)
          (10,0) (11,0) (12,0) (13,0) (14,0)
          (15,0) (16,0) (17,0) (18,0) (19,0)
        };

      \addplot[
        draw=black!55,
        line width=0.2pt,
        fill=ctiActor,
        forget plot
      ]
        coordinates {
          (0,0) (1,0) (2,0) (3,0) (4,0)
          (5,0) (6,0) (7,0) (8,0) (9,0)
          (10,0) (11,-1) (12,0) (13,0) (14,0)
          (15,0) (16,0) (17,0) (18,0) (19,0)
        };

      \addplot[
        draw=black!55,
        line width=0.2pt,
        fill=ctiDefence
      ]
        coordinates {
          (0,0) (1,0) (2,0) (3,0) (4,0)
          (5,0) (6,1) (7,10) (8,1) (9,0)
          (10,4) (11,7) (12,1) (13,2) (14,5)
          (15,10) (16,0) (17,2) (18,1) (19,35)
        };

      \addplot[
        draw=black!55,
        line width=0.2pt,
        fill=ctiDefence,
        forget plot
      ]
        coordinates {
          (0,0) (1,-1) (2,0) (3,0) (4,0)
          (5,0) (6,-6) (7,-11) (8,-2) (9,0)
          (10,-5) (11,-7) (12,-1) (13,0) (14,0)
          (15,-1) (16,0) (17,0) (18,0) (19,-3)
        };

      \addplot[
        draw=black!55,
        line width=0.2pt,
        fill=ctiVuln
      ]
        coordinates {
          (0,0) (1,0) (2,0) (3,0) (4,0)
          (5,0) (6,0) (7,0) (8,0) (9,0)
          (10,0) (11,0) (12,0) (13,0) (14,0)
          (15,0) (16,0) (17,0) (18,0) (19,0)
        };

      \addplot[
        draw=black!55,
        line width=0.2pt,
        fill=ctiVuln,
        forget plot
      ]
        coordinates {
          (0,0) (1,0) (2,0) (3,0) (4,0)
          (5,0) (6,0) (7,0) (8,0) (9,0)
          (10,0) (11,0) (12,0) (13,0) (14,0)
          (15,0) (16,0) (17,0) (18,0) (19,0)
        };

      \addplot[
        draw=black!55,
        line width=0.2pt,
        fill=ctiIoc
      ]
        coordinates {
          (0,0) (1,0) (2,0) (3,0) (4,0)
          (5,0) (6,1) (7,0) (8,0) (9,0)
          (10,0) (11,0) (12,0) (13,0) (14,0)
          (15,0) (16,0) (17,0) (18,1) (19,0)
        };

      \addplot[
        draw=black!55,
        line width=0.2pt,
        fill=ctiIoc,
        forget plot
      ]
        coordinates {
          (0,0) (1,0) (2,0) (3,0) (4,0)
          (5,0) (6,0) (7,-1) (8,0) (9,0)
          (10,0) (11,0) (12,0) (13,0) (14,0)
          (15,0) (16,0) (17,0) (18,0) (19,-1)
        };

      \legend{
        Technique,
        Tool,
        Malware,
        Infrastructure,
        Target,
        Actor,
        Defence,
        Vulnerability,
        Indicators
      }

      \draw[gray!70,dashed,thin]
        (axis cs:9.5,-44) -- (axis cs:9.5,57);

      \node[
        font=\footnotesize,
        anchor=south,
        text=black
      ]
        at (axis description cs:0.25,1.025)
        {Bot on \textit{Dread}};

      \node[
        font=\footnotesize,
        anchor=south,
        text=black
      ]
        at (axis description cs:0.75,1.025)
        {Bot on \textit{BHW}};

      \node[
        font=\footnotesize,
        rotate=-90,
        anchor=center,
        text=black
      ]
        at (axis description cs:1.045,0.72)
        {\SysName};

      \node[
        font=\footnotesize,
        rotate=-90,
        anchor=center,
        text=black
      ]
        at (axis description cs:1.045,0.22)
        {\textit{control}};

    \end{axis}
  \end{tikzpicture}

  \caption{
Per-topic CTI entities elicited within seven days of publication. Bars above the axis show entities in the thread assigned to \SysName; bars below show its matched control.}

  \label{fig:e3_pair_entities}
\end{figure}
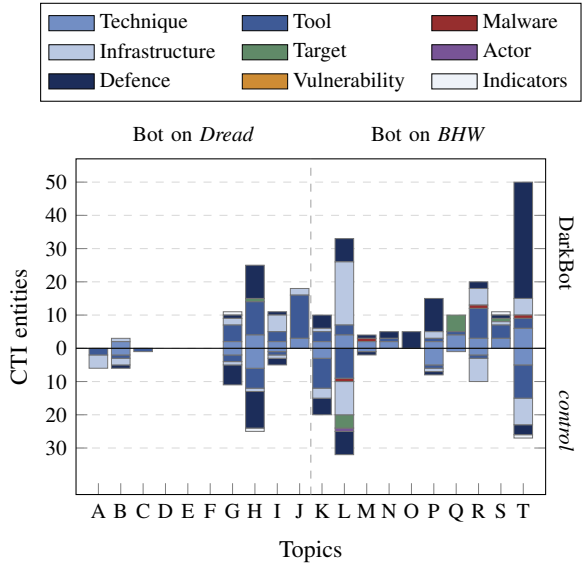

\mypari{Results}
Across the 20 completed pairs, threads to which \SysName was assigned carried 90 human messages containing a total of 223 CTI entities, against 84 on their matched controls with 154 entities over the same
seven-day windows. Bot-interacted threads achieved an average of 3.85 entities more than their matched control. 
The 95\% confidence interval for this mean difference ranges from 0.37 to 7.38 entities. An exact two-sided sign-flip permutation test yielded a $p$-value of 0.033, indicating that a difference at least this large would be unlikely under the null hypothesis of no effect.
Because assignment of the bot to one arm of each pair was counterbalanced rather than drawn independently, the test enumerates only the $\binom{20}{10} = 184{,}756$ assignments preserving the split between forums, which is the set of assignments that the randomization procedure could produce. The confidence interval is obtained by inverting the corresponding permutation test, so it excludes zero exactly when the $p$-value is below 0.05.
Figure~\ref{fig:e3_pair_entities} reports pair-level differences.
Beyond a higher count of CTI entities, some of the threads also produced concrete operational disclosures. For example,
in a matched thread on CAPTCHA bypass (Fig.~\ref{fig:ex_7}),  \SysName obtained details about how another user used Gemini to solve audio CAPTCHAs and mass-create GitHub accounts. This conversation has been reported to both Google and GitHub.  No comparable disclosure appeared in the matched control thread.
\begin{takeaway}
\textbf{Takeaway (EQ4).} Across 20 matched topic pairs, threads in which \SysName participated accumulated an average of 3.85 CTI entities more than their matched control. The 95\% confidence interval for this difference ranges from 0.37 to 7.38 entities, supporting a measurable positive effect on intelligence yield compared with standard passive approaches.
\end{takeaway}

\subsection{Automated Forum Deployment (EQ5)}
\label{sec:exp_real_forums}
Unlike the matched deployment in \S\ref{sec:exp_activity_e3}, this experiment evaluates
the full system on organically created threads discovered
and selected through its deployed engagement policy. Newly published candidate threads were screened by the \textsc{RelevanceAgent} before
interaction. 
We conducted 104 conversations across surface and dark web forums. On the surface web, this included 33 conversations on \textit{blackhatworld.com} (\textit{BHW}), 12 on \textit{sinister.ly}, and 2 on \textit{hackforums.net}. On the dark web, we carried out 57 conversations on \textit{Dread}~\cite{dread}.
Because interactions on \textit{sinister.ly} and \textit{hackforums.net} yielded only 2 responses in total, none of which provided relevant information, we focus our analysis on BlackHatWorld and Dread in the following.
Note that \textit{Dread} employs several interactive \texttt{CAPTCHA}s, which we bypassed manually for the scope of this research.

\mypar{Qualitative Interaction Analysis}
Live interactions demonstrated promising information-extraction capabilities, as the system often generated credible questions that elicited additional information about the topics discussed, particularly in the \textit{Dread} forum. Anonymized examples of interactions are provided in Appx~\ref{sec:app_examples}.

For instance, in the conversation regarding EDR evasion presented in Figure~\ref{fig:ex_1}, a user responded to the bot by stating that he had implemented and tested \textit{EDR-Freeze}~\cite{edrfreeze2025}, which had been published for the first time only one month before the conversation took place. In this case, the bot revealed additional information, indicating that the relatively new \textit{EDR-Freeze} technique was already being actively tested by potential threat actors.
Another relevant example is presented in Figure~\ref{fig:ex_4}, which discusses the trade-off between a data breach and other options. In this case, \SysName obtained detailed information about the structure of the stolen data and confirmation that the threat actor still had access to the compromised database. 
This information was not public before the interaction, showing that the system can surface CTI that would not emerge through passive monitoring alone.

However, in some cases, we observed reluctance among users to share details in public spaces. 
While it is not within the scope of this work to explore private conversations and interact with users, we nonetheless explore the private messages received as part of the CTI collected.
For instance, in one conversation, the bot successfully convinced a user to share the OSINT tools used in their operations, but the user preferred to do so in private chat, where the bot did not operate due to compliance with ethical prerequisites (see \S~\ref{sec:app_ethics}). 
After the deployment period, we manually inspected unsolicited messages received by the study accounts. In other cases, users shared additional information, external contact details, or service links through the forums' private
messaging systems, either to continue the exchange privately or to move it to another platform. 
In particular, these private messages included personal Telegram handles, Telegram links associated with services related to mail spoofing and large-scale scraping, and a Jabber address. This suggests that credible public participation enables access to other, more private channels and information, beyond increasing public activity and eliciting additional information in public threads.

\mypar{Threads Activity Analysis}
To characterize user activity in the organic deployment, we compare
bot-interacted threads with other non-interacted relevant threads from the
same forum sections and creation window. Unlike the matched
experiment in \S\ref{sec:exp_activity_e3},  the control threads topics may differ from the bot-interacted ones,
the results therefore describe an association and do not fully isolate completely
the causal effect of participation. Moreover, given the high number of threads involved, this evaluation uses message count as a proxy for user activity, which may not fully capture the CTI value of the resulting discussions.
We compare message activity across two groups of threads drawn from the same forum sections and the same creation time window. The first consists of all threads in which \SysName published at least one message and was re-fetched live after deployment to capture subsequent user activity, including any replies triggered by the bot. Bot messages are excluded from the count, so every counted message is a real user contribution. The second contains only threads from the same sections and window in which the bot did not interact, retaining only those classified as relevant by the \textsc{ThreadAnalyzer}, and \textsc{RelevanceAgent} pipeline, representing the baseline a passive monitor would observe. 

Figure~\ref{fig:comparison_passive} shows per-thread message distributions on both platforms. On Dread ($n_1=57$, $n_2=58$), bot-interacted threads have a median of 9 user messages, versus 6 for the passive baseline; on BHW ($n_1=33$, $n_2=45$), the medians are 18 and 7, respectively. These results indicate that active engagement by \SysName is associated with substantially richer activity than passive observation of the same communities.

\usepgfplotslibrary{statistics}
\usetikzlibrary{calc}

\pgfplotsset{
    box_passive/.style={
        thin, black,
        tick align=outside,
        major tick length=3pt,
        minor tick length=2pt,
        xtick pos=bottom,
        ytick pos=left,
        axis line style={thin},
        width=\columnwidth, height=2.8cm,
        xmin=-0.5, xmax=60,
        ymin=0, ymax=3,
        ytick={1,2},
        yticklabels={Bot-int., Others},
        yticklabel style={font=\scriptsize},
        xmajorgrids=true,
        grid style={dashed, gray!30, thin},
        xtick={0,10,20,30,40,50,60},
        tick label style={font=\scriptsize},
    },
    shared_legend_passive/.style={
        legend columns=-1,
        legend style={draw=none, font=\small, column sep=1.5ex},
        legend to name=sharedlegendpassive,
        legend image code/.code={
            \draw[##1, draw=none] (0cm,-0.1cm) rectangle (0.3cm,0.15cm);
        },
    }
}

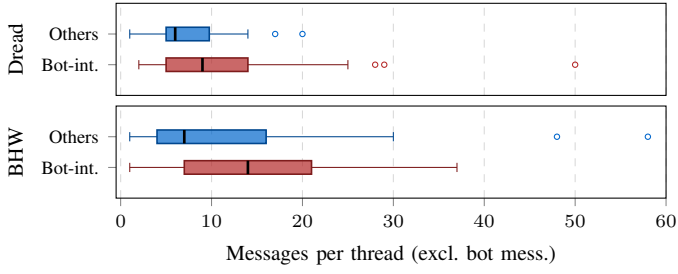
\begin{figure}[t]
\centering
\begin{tikzpicture}

    \begin{axis}[
        box_passive,
        name=top,
        ylabel={\footnotesize Dread},
        ylabel style={font=\footnotesize},
        xticklabels={},
        xtick style={draw=none},
    ]
        \fill[customred, fill opacity=0.7] (axis cs:5,0.78) rectangle (axis cs:14,1.22);
        \draw[customred!70!black, line width=0.6pt] (axis cs:5,0.78) rectangle (axis cs:14,1.22);
        \draw[black, line width=1pt] (axis cs:9,0.78) -- (axis cs:9,1.22);
        \draw[customred!70!black] (axis cs:2,1) -- (axis cs:5,1);
        \draw[customred!70!black] (axis cs:2,0.85) -- (axis cs:2,1.15);
        \draw[customred!70!black] (axis cs:14,1) -- (axis cs:25,1);
        \draw[customred!70!black] (axis cs:25,0.85) -- (axis cs:25,1.15);

        \fill[customblue, fill opacity=0.7] (axis cs:5,1.78) rectangle (axis cs:9.75,2.22);
        \draw[customblue!70!black, line width=0.6pt] (axis cs:5,1.78) rectangle (axis cs:9.75,2.22);
        \draw[black, line width=1pt] (axis cs:6,1.78) -- (axis cs:6,2.22);
        \draw[customblue!70!black] (axis cs:1,2) -- (axis cs:5,2);
        \draw[customblue!70!black] (axis cs:1,1.85) -- (axis cs:1,2.15);
        \draw[customblue!70!black] (axis cs:9.75,2) -- (axis cs:14,2);
        \draw[customblue!70!black] (axis cs:14,1.85) -- (axis cs:14,2.15);

        \addplot[only marks, mark=o, mark size=1pt, color=customred]
            table[y=x, x=y, col sep=comma,
                  restrict expr to domain={\thisrow{x}}{0.5:1.5}]
            {Plots/fixed_plot/dread_outliers.csv};
        \addplot[only marks, mark=o, mark size=1pt, color=customblue]
            table[y=x, x=y, col sep=comma,
                  restrict expr to domain={\thisrow{x}}{1.5:2.5}]
            {Plots/fixed_plot/dread_outliers.csv};
    \end{axis}

    \begin{axis}[
        box_passive,
        name=bottom,
        at={(top.south)}, anchor=north, yshift=-4pt,
        xlabel={\footnotesize Messages per thread (excl.\ bot mess.)},
        ylabel={\footnotesize BHW},
        ylabel style={font=\footnotesize},
        shared_legend_passive,
    ]
        \addlegendimage{fill=customblue, fill opacity=0.6, draw=none}
        \addlegendentry{Other threads}
        \addlegendimage{fill=customred, fill opacity=0.7, draw=none}
        \addlegendentry{Bot-interacted}

        \fill[customred, fill opacity=0.7] (axis cs:7,0.78) rectangle (axis cs:21,1.22);
        \draw[customred!70!black, line width=0.6pt] (axis cs:7,0.78) rectangle (axis cs:21,1.22);
        \draw[black, line width=1pt] (axis cs:14,0.78) -- (axis cs:14,1.22);
        \draw[customred!70!black] (axis cs:1,1) -- (axis cs:7,1);
        \draw[customred!70!black] (axis cs:1,0.85) -- (axis cs:1,1.15);
        \draw[customred!70!black] (axis cs:21,1) -- (axis cs:37,1);
        \draw[customred!70!black] (axis cs:37,0.85) -- (axis cs:37,1.15);

        \fill[customblue, fill opacity=0.7] (axis cs:4,1.78) rectangle (axis cs:16,2.22);
        \draw[customblue!70!black, line width=0.6pt] (axis cs:4,1.78) rectangle (axis cs:16,2.22);
        \draw[black, line width=1pt] (axis cs:7,1.78) -- (axis cs:7,2.22);
        \draw[customblue!70!black] (axis cs:1,2) -- (axis cs:4,2);
        \draw[customblue!70!black] (axis cs:1,1.85) -- (axis cs:1,2.15);
        \draw[customblue!70!black] (axis cs:16,2) -- (axis cs:30,2);
        \draw[customblue!70!black] (axis cs:30,1.85) -- (axis cs:30,2.15);

        \addplot[only marks, mark=o, mark size=1pt, color=customblue]
            table[y=x, x=y, col sep=comma]
            {Plots/fixed_plot/bhw_outliers.csv};
    \end{axis}

\end{tikzpicture}
\caption{User messages per thread across bot-interacted and non-interacted relevant threads during the time window.}
\label{fig:comparison_passive}
\end{figure}

\mypar{Responses Analysis}
On \textit{BHW}, \SysName published 103 questions across 33 threads and received 112 qualifying user replies. We define a qualifying user reply as a message from another user account, posted after \SysName began interacting, that the pipeline identified as responding to \SysName; unrelated messages are excluded. This corresponds to 1.09 replies per question and 3.39 replies per thread. Of the 33 threads, 22 (66.7\%) received at least one qualifying reply, with 5.09 replies per responding thread.
On Dread, \SysName published 95 questions across 57 threads and received 42 qualifying user replies, corresponding to 0.44 replies per question and 0.74 replies per thread. Of the 57 threads, 16 (28.1\%) received at least one qualifying reply, with 2.63 replies per responding thread. Although response density was lower than on BHW, the Dread conversations generally focused more directly on CTI-relevant topics.

\mypari{Message length} 
On \textit{BHW}, user replies were substantially longer than system
messages on average (221.5 compared to 130.0 characters), suggesting that the system’s questions elicited more elaborate responses. 
Conversely, on \textit{Dread}, message lengths were more balanced (141.1 characters for user replies, 152.4 for system messages). This reflects a more concise communication style typical of \textit{Dread} users, where interactions tend to be direct.  

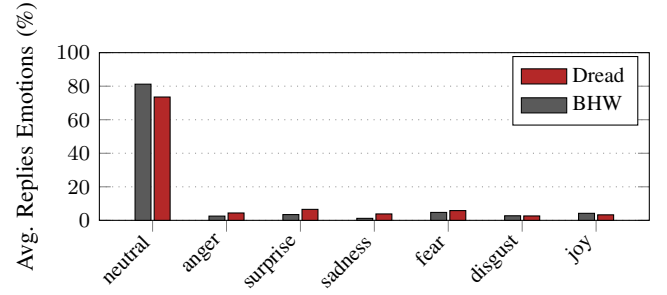
\begin{figure}[t]
\centering
\begin{tikzpicture}
\begin{axis}[
    ybar,
    bar width=6pt,
    width=\linewidth,
    height=3.8cm,
    ymin=0,
    ymax=100,
    ylabel={Avg. Replies Emotions (\%)},
    enlarge x limits=0.12,
    symbolic x coords={neutral,anger,surprise,sadness,fear,disgust,joy},
    xtick=data,
    x tick label style={rotate=45, anchor=east, font=\footnotesize},
    tick label style={font=\footnotesize},
    label style={font=\small},
    xtick pos=left,
    ymajorgrids=true,
    grid style={dotted, gray!80},
legend style={
    font=\footnotesize,
    at={(0.98,0.98)},
    anchor=north east,
    legend cell align={left},
    reverse legend,
},
    area legend,
]
\addplot[draw=black, fill=customgrey, bar shift=-3.6pt] coordinates {
    (neutral,81.20) (anger,2.53) (surprise,3.44) (sadness,1.19) (fear,4.73) (disgust,2.72) (joy,4.19)
};
\addlegendentry{BHW}
\addplot[draw=black, fill=customred, bar shift=3.6pt] coordinates {
    (neutral,73.54) (anger,4.39) (surprise,6.55) (sadness,3.82) (fear,5.81) (disgust,2.62) (joy,3.28)
};
\addlegendentry{Dread}
\end{axis}
\end{tikzpicture}
\caption{Emotion distribution of received replies}
\label{fig:emotion_distribution_grouped}
\end{figure}

\mypari{Emotion Distribution}
We analyzed emotions using Hartmann's English emotion classification model~\cite{hartmann2022emotionenglish}. 
As shown in Fig.~\ref{fig:emotion_distribution_grouped}, \textit{Neutral} tones dominate across both platforms, with negative sentiment remaining limited and never escalating into adversarial interactions. On the surface web, this aligns with discussions centered on SEO, services, and technical advice. On the dark web, while neutrality remains dominant, higher proportions of \textit{Anger} and \textit{Fear} were observed, particularly in discussions involving exploits, fraud, and operational security.

\mypari{Timing Analysis}
Deployments spanned approximately three weeks on \textit{BHW} and one month on \textit{Dread}.
As shown in Figure~\ref{fig:timing_violin}, \textit{BHW}
exhibited considerably faster dynamics: the median time to the first received reply is 2.8~h compared to 7.2~h on \textit{Dread}, and median conversation durations are roughly half as long (48.5~h vs.\ 82.8~h). The conversation duration is defined as the time elapsed between the bot's first message and the last message in the thread that received no further replies.
Appx~\ref{sec:app_week_timing} also suggests that weekly interaction patterns further differentiate the two platforms.

\input{Plots/timing_violin}

\mypar{Stylistic Credibility Analysis}
\label{sec:exp_real_style}
Across all monitored interactions (\S\ref{sec:exp_activity_e3}, \S\ref{sec:exp_real_forums}), we observed no account suspensions, moderator interventions, or explicit accusations that \SysName was automated; its messages did not lead to any visible moderation or enforcement action. During the same observation period, we also observed other accounts exhibiting overtly automated posting behavior being
identified and suspended by forum moderators. Although this does not constitute a controlled comparison, it indicates that
moderation and bot-related enforcement were active on the studied platforms.

\begin{figure}[t]
\centering
\begin{tikzpicture}
\begin{axis}[
    width=\columnwidth,
    height=4cm,
    ymin=-0.6, ymax=1.6,
    xmin=-0.05, xmax=1.05,
    ytick={0,1},
    yticklabels={Dread, BHW},
    yticklabel style={font=\small},
    xlabel={Avg. AI Detection Score},
    xlabel style={font=\small},
    xticklabel style={font=\small},
    xtick={0, 0.2, 0.4, 0.6, 0.8, 1.0},
    xmajorgrids=true,
    ymajorgrids=false,
    grid style={dotted, gray!80},
    clip=false,
    legend style={
        at={(0.5,1.12)},
        anchor=south,
        draw=none,
        font=\small,
        legend columns=3,
        column sep=0.5em,
        /tikz/every odd column/.append style={yshift=0.4pt},
    },
]

\addplot[
    only marks,
    mark=*,
    mark size=1.5pt,
    fill=customgreen,
    draw=customgreen,
    fill opacity=0.5,
] coordinates {
    (0.00,1) (0.00,1) (0.00,1) (0.00,1) (0.00,1)
    (0.80,1) (1.00,1) (1.00,1) (1.00,1) (0.96,1)
    (0.00,1) (0.00,1) (0.00,1) (0.00,1) (0.00,1)
    (0.91,1) (0.00,1) (0.00,1) (0.00,1) (0.06,1)
    (0.00,1) (1.00,1) (0.99,1) (0.00,1) (0.00,1)
    (0.09,1) (0.03,1) (0.75,1) (0.04,1) (0.00,1)
    (0.52,1) (0.00,1) (0.67,1) (0.17,1) (0.01,1)
    (0.00,1) (0.00,1) (0.00,1) (0.00,1) (1.00,1)
    (0.00,1) (0.59,1) (0.00,1) (1.00,1) (0.14,1)
    (0.06,1) (0.00,1) (0.00,1) (0.00,1) (0.00,1)
    (0.00,1) (0.00,1) (0.40,1) (0.00,1) (0.00,1)
    (0.00,1) (0.00,1) (0.00,1) (0.99,1) (0.00,1)
    (0.97,1) (0.00,1) (1.00,1) (1.00,1) (0.00,1)
    (0.67,1) (0.00,1) (0.00,1) (0.00,1) (0.00,1)
    (0.01,1) (0.00,1) (0.34,1) (0.88,1)
};
\addlegendentry{Users}

\addplot[
    only marks,
    mark=diamond*,
    mark size=4pt,
    fill=customred,
    draw=customred,
] coordinates { (0.07, 1) };
\addlegendentry{\SysName}

\addplot[
    only marks,
    mark=|,
    mark size=5pt,
    very thick,
    draw=black,
] coordinates { (0.2758, 1) };
\addlegendentry{User mean}

\addplot[
    only marks,
    mark=*,
    mark size=1.5pt,
    fill=customgreen,
    draw=customgreen,
    fill opacity=0.5,
] coordinates {
    (0.00,0) (0.10,0) (0.11,0) (0.02,0) (0.00,0) (0.00,0)
    (0.00,0) (0.00,0) (0.00,0) (0.00,0) (0.00,0)
    (0.00,0) (0.33,0) (0.00,0) (0.00,0) (0.29,0)
    (0.00,0) (0.00,0) (0.00,0) (0.00,0) (0.00,0)
    (0.00,0) (0.19,0) (0.13,0) (0.00,0) (0.01,0)
    (0.00,0) (1.00,0) (0.00,0) (0.00,0) (0.17,0)
    (0.00,0) (0.81,0) (0.00,0) (0.20,0) (0.00,0)
    (0.00,0) (0.00,0) (0.50,0) (0.88,0) (0.00,0)
    (0.00,0) (0.00,0) (0.00,0) (0.00,0) (0.01,0)
    (0.00,0) (0.51,0) (0.14,0) (0.00,0) (0.00,0)
    (0.00,0) (0.02,0) (0.00,0) (0.00,0) (0.00,0)
    (0.00,0) (0.07,0) (0.00,0) (0.38,0)
};

\addplot[
    only marks,
    mark=diamond*,
    mark size=4pt,
    fill=customred,
    draw=customred,
] coordinates { (0.1075, 0) };

\addplot[
    only marks,
    mark=|,
    mark size=5pt,
    very thick,
    draw=black,
] coordinates { (0.1309, 0) };

\end{axis}
\end{tikzpicture}
\caption{AI detection scores per user on BHW and Dread.
}
\label{fig:ai-detection-per-user}
\end{figure}
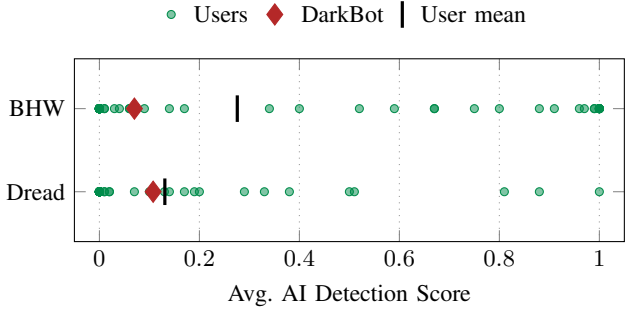

\mypari{AI-detection Analysis}
We evaluated the generated messages using the Sapling API~\cite{sapling}, considered among the most reliable freely available AI detectors~\cite{agarwal2025ai_content_detector}.
\SysName messages averaged detection scores of 6.87\% on \textit{BHW} and 10.75\% on \textit{Dread}, compared with 27.58\% and 13.09\%, respectively, across other forum users.
Figure~\ref{fig:ai-detection-per-user} shows the average score per user.
These results suggest that the detector does not distinguish \SysName from typical user content, supporting the pipeline's ability to conceal its automated nature. Notably, \textit{BHW} users received higher scores than dark-web users, possibly reflecting automation or AI-assisted participation. Appx~\ref{app:syle_embeddings}
provides an additional qualitative style-embedding analysis.

\begin{takeaway}
  \textbf{Takeaway (EQ5).} Across 104 live conversations and 213 published messages, the system never triggered any account suspension, moderator intervention, or explicit AI accusation. 
  \SysName successfully elicited CTI that would not have been accessible through passive monitoring, including newly published attack techniques already being tested in the wild and details about active database breaches.
Notably, some users preferred to continue the exchange
through the forums' private chats, where they shared
additional information, further demonstrating how credible engagement can
surface content unavailable to passive monitoring of public
threads.
\end{takeaway}

\section{Threats to Validity}
\label{sec:threats_to_validity}

While our results suggest that multi-agent architectures can create credible bots that elicit \ac{CTI}-related information from underground communities, several threats to validity remain. We discuss them as internal and external validity threats.

\mypar{Internal Threats}
The intelligence probing evaluation (\S\ref{sec:exp_extraction}) depends on a simulated user whose behavior may not fully reflect real adversarial dynamics. 
Moreover, the ground truth for techniques is manually validated. This is because the extraction of TTPs from unstructured forum content remains a difficult problem with limited state-of-the-art performance, and the \textsc{TTP Extractor} produces a non-negligible number of false positives.
This choice reduces the closed-loop dependence on the extractor but
introduces annotator judgment, particularly for implicit or ambiguous ATT\&CK mappings. 
On the other hand, highly relevant questions may be penalized when the corresponding topic was not discussed in the original CrimeBB conversation, leading to an underestimation of the system's elicitation capability.
The robustness evaluation (\S\ref{sec:exp_robustness}) uses general-purpose jailbreak benchmarks. Design-aware adversaries crafting system-specific manipulations could be more effective than the tested prompts.

\mypar{External Threats}
The survey-based credibility evaluation (\S\ref{sec:exp_style_survey}) relies on participants with a computer science background, who may, however,
lack familiarity with the linguistic conventions and slang typical of underground forums.
The live deployment is limited to specific forums (\S\ref{sec:exp_real_forums}), however, the effectiveness of this approach might vary across different online platforms and communities with different norms and moderation practices. Additionally, underground forums are subject to significant linguistic and topic drift over time. 
Ethical minimization requirements limited the scale and duration of the live deployment (\S~\ref{sec:app_ethics}), reducing statistical power and generalizability. Exact replication is also difficult because forum content, moderation policies, and user behavior may change over time, especially after study disclosure. Although we release links to the public evaluation threads, online content may later change, be removed, or migrate.

\section{Conclusion}
This paper presented \SysName, a multi-agent \ac{LLM}-based system designed to actively elicit cyber threat intelligence from public underground forums through targeted conversational engagement. Unlike prior approaches that rely on passive crawling or single-agent designs limited to specific tasks, \SysName addresses the broader goal of \ac{CTI} elicitation by decomposing the interaction task into 11 specialized agents that handle engagement gating, question generation, and linguistic style adaptation.
Our evaluation showed that the system can recover 72.8\% of MITRE ATT\&CK techniques in a simulated setting from the original CrimeBB conversations, starting only from the initial post, confirming its ability to understand context and generate coherent questions.
The system effectively detects harmful content and is robust against adversarial inputs. The pipeline fully rejected all tested jailbreak prompts and filtered the majority of content that violated policies, with the overall architecture compensating for the limitations of individual components. The multi-agent design outperformed a single-agent baseline in  tactical coverage, and stylistic credibility. In a live prospective experiment with 20 matched topic pairs, threads assigned to \SysName accumulated an average increase of 3.85 CTI entities over seven days. 
Across 104 live forum threads, the deployment triggered no account bans, moderator interventions, or AI accusations, while eliciting \ac{CTI}-relevant information unavailable through passive crawling.

\mypar{Future work} 
\label{sec:future}
Future work includes extending the pipeline to multilingual forums and broader platforms, developing adaptive elicitation strategies based on community feedback, and conducting longitudinal deployments to assess sustained detection risk. 
Further directions include targeting emerging threats (e.g., new CVEs or threat actors), supporting conversation migration to private channels, and automating discovery of hidden underground communities.

\section*{Acknowledgments}
We thank Professor Gianluca Stringhini for his valuable feedback and helpful suggestions. We also thank the Cambridge Cybercrime Centre for granting us access to the CrimeBB dataset. This work was supported by the Ministero delle Imprese e del Made in Italy as part of the project ``WASABI - Web Analysis System with Artificial intelligence on BIg data'' under the Innovation Agreement CUP B49J25000080005 and by BV TECH S.p.A.


\section{Ethical Considerations}
This work involves automated interaction with online communities that may discuss or be involved in harmful or illegal activities, and the development of techniques enabling automated human-mimicking conversational behavior.
As such, the study raises ethical concerns related to the risk of enabling or facilitating criminal behavior, the safety of interactions, and the potential misuse of the proposed technology beyond its intended defensive purpose. 
To address these concerns, the research has undergone institutional ethical review and data protection impact assessment, with approval from the corresponding institutional authorities. 
The proposed system design incorporates safety guardrails and content filters and follows a data minimization approach.
Furthermore, the \ac{LLM} models used in this research are self-hosted to ensure that all data and information remain within the controlled research environment, preventing any potential transmission of sensitive or harmful data. 
All procedures adhere to applicable ethical standards for human-subjects research and to data protection regulations, including the principles of data minimization, purpose limitation, and proportionality.
\label{sec:app_ethics}
\subsection{Institutional Evaluation Process}
Before starting the research activity, the study underwent an institutional evaluation process that involved both a privacy compliance review and an ethical assessment by the university's Research Ethics Committee.
The research team prepared the required documentation, including: (i) a data processing description form detailing the nature of the data collected, the legal basis for processing, and the data protection measures adopted; (ii) a Data Protection Impact Assessment (DPIA) analyzing the risks associated with the processing and the corresponding mitigation measures; (iii) a formal request for ethical review by the Research Ethics Committee; and (iv) a declaration of commitment by the principal investigator to enforce non-disclosure agreements on the collected data with all personnel involved in the project.
The documentation was first reviewed by the departmental Privacy Officer, who coordinated with the institutional Data Protection Officer (DPO). Both issued a favorable opinion, confirming that the processing is lawful, proportionate, and compliant with GDPR principles.
Following the privacy approval, the complete documentation was submitted to the institutional Research Ethics Committee. The research team attended a dedicated hearing with the Committee, during which the study's methodology, ethical implications, and safeguards were discussed in detail. The Committee granted approval to proceed with the research.
\subsection{Human Subjects and Consent}
The research involves indirect interaction with human participants who post content on publicly accessible forums under pseudonyms. Obtaining individual informed consent is impractical and would compromise the scientific validity of the study, as one of the research goals is to evaluate whether the conversational agent can operate indistinguishably from a human user. In line with established ethical guidelines for observational research in public online spaces, the study is limited to forums that are openly accessible and does not attempt to identify, contact, or profile individual users. To ensure transparency despite the absence of individual notice, a general privacy notice describing the research objectives, categories of data processed, legal basis, and available data subject rights has been published on the institution’s official website, in accordance with applicable data protection regulations.

\subsection{Personal Data and Privacy}
While no directly identifying information (e.g., real names, email addresses, or physical addresses) is collected, forum pseudonyms and textual content may theoretically allow re-identification when combined with external data. Following a precautionary approach, all collected data are treated as personal data under the GDPR. Usernames are immediately pseudonymized using salted hashing, data are stored exclusively on secure institutional infrastructure, and access is restricted to authorized researchers. Raw data are retained only for the minimum time necessary for analysis and are permanently deleted thereafter, with only aggregated and anonymized results preserved.

\subsection{Deception and Interaction Risks}
The conversational agent intentionally does not disclose its automated nature in order to preserve experimental validity. This form of limited deception is carefully constrained: the agent does not impersonate real individuals or organizations, and does not encourage or assist in illegal activities. The agent is restricted to asking clarifying or exploratory questions related to cybersecurity topics and is equipped with safeguards to detect and terminate interactions involving sensitive domains (e.g., violence, drugs, weapons, hate, health, or judicial information), as well as attempts at jailbreaking or eliciting malicious assistance.

The study is designed to minimize risks to participants. Interactions occur in environments where users already expect public discussion with unknown pseudonymous peers, and the agent’s behavior is designed to align with typical conversational norms
and interactional conventions observed in the target
communities. If the agent is suspected of being automated or if unexpected interaction patterns emerge, the conversation is automatically terminated. Moreover, experiments requiring live forum interactions were manually monitored by researchers, even if no intervention was ultimately needed. The research does not involve interventions that could reasonably cause psychological distress, harm, or material risk to participants.

\subsection{Dual-Use and Misuse Concerns}
\label{sec:app_policies}
The techniques explored in this work enable automated, human-like interaction, and could potentially be misused for unethical purposes such as large-scale social engineering, surveillance, or disinformation campaigns.  To mitigate these risks, the study focuses on defensive cybersecurity applications and restricts access to the \SysName codebase to scientific use only, subject to authors' approval, requiring the submission of a detailed request form describing the intended use and an ethical justification.
The intended contribution of this research is to improve threat intelligence capabilities and the understanding of emerging cyber threats, thereby supporting defensive and preventive security efforts.
\subsection{Engagement Guidelines}
To mitigate potential harms, we defined a set of engagement guidelines that constrain the system's behavior during live interactions.
The system exclusively engages in public forum threads, where users are aware that their messages are publicly visible. We avoid automated interaction and data collection from private chats, although results suggest they may yield richer CTI (\S\ref{sec:exp_real_forums}). Forums requiring payment, invitation, or proof of affiliation to access were excluded, as operating in such spaces would imply a higher degree of infiltration and a greater risk of harm to the community.
The system never downloads attachments or external files to avoid inadvertently obtaining or distributing harmful content such as malware or pornographic images.
The system is explicitly prohibited from engaging in discussions involving: drugs and illegal substances, weapons, pornographic or adult material, violence, and any content involving judicial, medical, or health-related personal information. These restrictions were enforced to avoid eliciting sensitive information that could directly harm individuals or facilitate illegal activities beyond the scope of CTI collection. In cases where live interactions surface evidence of ongoing or imminent harm (e.g., active data breaches or active exploitation of individuals), such findings are reported to the relevant parties following standard responsible disclosure practices.

\bibliographystyle{IEEEtran}
\bibliography{bibliography}

\section*{Generative AI Usage}
The following generative AI tools were used during the preparation of this paper. Claude (Anthropic) was used to assist in drafting and refining portions of the written text, to assist in the development of experimental code, and to produce drafts of LaTeX code for data visualizations; all AI-assisted writing was reviewed and revised by the authors, code was manually validated, and LaTeX visualization code was manually checked by cross-checking the resulting plots against the underlying experimental data. TrinkAI/WriteFull and Wordtune were used for grammar checking and paraphrasing or condensing portions of the text. All AI-assisted content was reviewed by the authors to ensure accuracy and consistency with the research.

\appendices
\section{Open Science}
\label{sec:app_open}

In the interest of transparency and reproducibility, we describe below the availability of the artifacts associated with this work.
\footnote{Due to ethical considerations, the artifact will be made available only upon request}

\mypar{Code} The \SysName codebase, including all agent prompts, orchestration logic, and evaluation scripts, will be available upon request. Access is granted subject to a brief description of the intended use and a related ethical discussion. Requests should be directed to the corresponding author.

\mypar{Models} All base models used in this work are publicly available open-weight models (e.g., Gemma3-27B, Qwen3-32B). Full model details and references are provided in Appx~\ref{sec:app_implementation}. The fine-tuned \textsc{SlangConverter} checkpoint will be released alongside the codebase under the same access conditions.

\mypar{Datasets}  
The AnnoCTR, BeaverTails, CoDA, and jailbreak datasets used in evaluation are publicly available, and links to their sources are provided in the references.
The CrimeBB dataset used for training and evaluation is governed by a data sharing agreement and cannot be redistributed.  The synthetic slang conversion dataset generated for fine-tuning the \textsc{SlangConverter} cannot be released, as it is derived from CrimeBB data and subject to the same data protection constraints. Although the datasets
themselves cannot be redistributed, the codebase includes details on how to extract the subsets employed in our experiments, so that researchers who obtain access to the original datasets can reproduce our experimental conditions exactly. URLs of the public threads involved in the live experiments (\S\ref{sec:exp_activity_e3}, \S\ref{sec:exp_real_forums}) are provided with the code. Note that the URLs reference public forums and threads that are not under our control and may evolve, be migrated or removed. Details on datasets and the derivation of subsets are reported in Appx~\ref{sec:app_datasets}.

\section{Implementation Details}
\label{sec:app_implementation}
We implemented the multi-agent pipeline using Microsoft AutoGen~\cite{autogen},
which provides the inter-agent communication primitives and orchestration logic.
Since conversations may span multiple days, we use Temporal.io~\cite{temporal}
to handle workflow persistence and automatic retries on failure.
Table~\ref{tab:models} reports the model assigned to each agent. To comply with ethical guidelines, we chose open-weight models running on offline hardware rather than relying on commercial models. This avoids sharing potentially sensitive personal information with model providers. Agents requiring uncensored outputs, specifically those that must reason about attack behaviors
without triggering safety refusals, use an abliterated variant of Gemma~3~27B~\cite{gemma_2025}.
Agents responsible for logical reasoning and classification, which do not require obliteration, use Qwen3-32B~\cite{qwen3technicalreport}. 
The \textsc{SlangConverter} uses a fine-tuned 4B-parameter model, as the conversion task is narrow and does not require large-scale reasoning capacity. In particular, it was fine-tuned using Unsloth~\cite{unsloth} on a synthetic dataset derived from CrimeBB~\cite{crimebb}. The dataset was constructed by pairing real CrimeBB user messages containing slang with their formal rewrites, generated by an LLM (\textsc{SyntheticAgent}). The \textsc{SlangConverter} was then trained on this dataset to perform the inverse conversion: given a formally generated question, it produces a slang version that incorporates the specific terminology and stylistic patterns of real dark web users. The lower section of Table~\ref{tab:models} reports support agents, which are not included in the pipeline. Specifically, the \textsc{FakeUserAgent} was employed in specific settings of the experimental validation (presented in Section~\ref{sec:exp_extraction}), and the \textsc{SyntheticAgent} was employed to generate the synthetic dataset used to fine-tune the \textsc{SlangConverter}.
We ran the models on an Ollama server hosted in an Ubuntu 24.04 LTS virtual machine with an Intel(R) Xeon(R) Gold 6418H 2.100GHz, 64GB of RAM, and an NVIDIA L40 GPU with 48GB of VRAM.

\begin{table}[h]
    \caption{Model Selection for Each Agent}
    \label{tab:models}
  \begin{tabular}{ccc}
    \toprule
    Agent&Model&\\
    \midrule
    \textsc{ThreadAnalyzer} & gemma3-abliterated-GGUF-q4-27b&\cite{labonne2025gemma3abliterated} \\
    \textsc{RelevanceAgent} & qwen3-32b&\cite{qwen3technicalreport} \\
    \textsc{ToneAgent} & qwen3-32b&\cite{qwen3technicalreport} \\
    \textsc{TTP Extractor} & gemma3-abliterated-GGUF-q4-27b&\cite{labonne2025gemma3abliterated} \\
    \textsc{TacticSelector} & qwen3-32b&\cite{qwen3technicalreport} \\
    \textsc{QuestionAgent} & gemma3-abliterated-GGUF-q4-27b&\cite{labonne2025gemma3abliterated} \\
    \textsc{ReviewerAgent} & gemma3-abliterated-GGUF-q4-27b&\cite{labonne2025gemma3abliterated} \\
    \textsc{ReplyAnalyzer} & qwen3-32b&\cite{qwen3technicalreport} \\
    \textsc{DefensiveAgent} & qwen3-32b&\cite{qwen3technicalreport} \\
    \textsc{FilterAgent} & qwen3-32b&\cite{qwen3technicalreport} \\
    \textsc{SlangConverter} & gemma3-4b-unsloth-bnb-4bit $^\star$ &\cite{gemma_2025,unsloth}\\
    \midrule
    \textsc{FakeUserAgent} & qwen3-32b&\cite{qwen3technicalreport} \\
    \textsc{SyntheticAgent} & gemma3-abliterated-GGUF-q4-27b&\cite{labonne2025gemma3abliterated} \\
  \bottomrule
  \\
  \multicolumn{2}{c}{$^\star$: the model has been fine-tuned for its specific task.} 
\end{tabular}
\end{table}

\section{Stylistic Credibility Survey}
The survey was administered via Google Forms and distributed to participants with a Computer Science background through universities and social networks, without any form of compensation or incentive. After a consent form explaining the nature of the study and warning that some content may contain offensive or vulgar language, participants were presented with a brief description of the task and a visual example illustrating the response format.
The survey was divided into two sections. The first section, \textit{Style and Credibility}, asked participants to judge which of two replies seemed more likely to have been written by a human underground forum user, with a third option, "Equally Likely" available. The second section, \textit{Tone Alignment}, asked participants which reply better followed the conversational tone of the thread (e.g., informal, paranoid, aggressive) and better mimicked its stylistic patterns (e.g., slang, typos, emojis, abbreviations), with "Equally Well" as the neutral option.
Five representative threads were selected from CrimeBB~\cite{crimebb}, covering topics such as DDoS, botnet setup, bulletproof hosting, and script marketplaces. The \textit{Style and Credibility} section presented two comparison conditions: the full \SysName pipeline versus the single-agent baseline, and \SysName versus the pipeline without the \textsc{SlangConverter}. The \textit{Tone Alignment} section presented the third condition: \SysName versus the pipeline without the \textsc{ToneAgent}. The assignment of replies to positions A and B was randomized to avoid position bias. The survey collected 118 valid responses.
\label{sec:app_survey}

\section{Datasets Details}
\label{sec:app_datasets}
\mypar{CoDA}
The dataset~\cite{coda} contains 10{,}000 dark web documents labeled into ten topic categories. The data set is not specifically designed for the assessment of \ac{CTI}-relevance identification, and the labels just represent the main high-level topic of the document. However, to the best of the authors' knowledge, a dataset specific to this task containing dark web forum conversations is missing, and therefore, the CoDA dataset is the best option for validating the relevance identification logic.
For computational reasons, we considered only a subset of the dataset, consisting of 470 texts. Specifically, we selected the first 30 samples from each category except for the hacking category, from which we drew 200 samples given its stronger relevance to \ac{CTI}. We removed duplicate entries and excluded texts exceeding the model context window (15,000 tokens). This dataset is used to evaluate engagement gating and robustness against prohibited topics (EQ2, §\ref{sec:exp_robustness}).

\mypar{CrimeBB}
The dataset~\cite{crimebb} is actively maintained and periodically expanded, containing over 99 million posts from 34 forums in different languages. We selected a subset of 100 relevant English threads from the September 2024 snapshot.
We selected conversations containing between 15 and 150 messages and including only those containing one or more keywords belonging to a predefined set of relevant terms (e.g., exploit, malware, DDoS, antivirus, 0day). Once the filtering was applied, we manually reviewed each conversation to make sure it contained relevant discussions (e.g., removing threads consisting only of one relevant message and several acknowledgments). 
This data set is used in the information extraction evaluation and design validation experiments (EQ1 and EQ3, §\ref{sec:exp_extraction} and §\ref{sec:exp_design}), and to construct the training data for the \textsc{SlangConverter}.

\mypar{AnnoCTR} The dataset~\cite{annoctr} contains cyber threat reports from commercial \ac{CTI} vendors, annotated by domain experts with tactics and techniques from the MITRE ATT\&CK framework, covering both explicit and implicit mentions.
We use this dataset as ground truth to evaluate the \textsc{TTP Extractor}'s ability to identify ATT\&CK tactics and techniques. For our evaluation, we considered only the 105 reports that contain at least one ground-truth MITRE ATT\&CK Technique annotation. This dataset is used to evaluate the TTP Extractor against labeled ground truth (EQ3, §\ref{sec:exp_design}).

\mypar{Jailbreak Prompts}
We tested the system using two publicly available jailbreak datasets:
\textit{Do Anything Now} by Shen et al.~\cite{do_anything_now}, from which we sampled the first 10\% of the adversarial evaluation set (restricted to
\textit{data\_type=adversarial\_harmful}), and \textit{In-the-Wild Jailbreaks}
by Jiang et al.~\cite{in_the_wild}, using the \textit{jailbreak\_2023\_05\_07}
subset. After removing duplicates, the resulting dataset consists of a total of 845 jailbreak prompts.
This dataset is used to evaluate robustness against adversarial replies and pipeline-level safety (EQ2, §\ref{sec:exp_robustness}).

\mypar{BeaverTails} The dataset~\cite{question_answer_harmful_topics} contains question-answer pairs annotated for harmful content across several categories. We use a subset of questions belonging to the following harmful categories: drug\_abuse, weapons, banned\_substances, self\_harm, sexually\_explicit, adult\_content, terrorism, and organized\_crime. This subset is used to evaluate the robustness of the final message filter against harmful content that may have bypassed earlier safety components (EQ2, §\ref{sec:exp_robustness}).

\mypar{Slang Conversion Dataset}
The dataset was constructed by pairing real CrimeBB user messages containing slang with their formal rewrites, generated by an LLM (\textsc{SyntheticAgent}). The \textsc{SlangConverter} was then trained on this dataset to perform the inverse conversion: given a formally generated question, it produces a slang version that incorporates the specific terminology and stylistic patterns of real dark web users. This dataset is used exclusively for fine-tuning the SlangConverter.

\section{LLamaGuard3 Comparison}
\label{sec:app_llamaguard}
The \textsc{RelevanceAgent}, \textsc{DefensiveAgent}, and \textsc{FilterAgent} described in §\ref{sec:exp_design} perform the task of identifying prohibited topics and jailbreak attempts using a general-purpose reasoning \ac{LLM} equipped with a structured prompt that encodes exclusion rules and positive \ac{CTI} indicators.
To quantify the advantage of this approach over simpler alternatives, we evaluate
LLaMA Guard~3~8B~\cite{dubey2024llama3herdmodels} as a single-model baseline on the same CoDA subset
described in Appx~\ref{sec:app_datasets}, comparing it with the performances of the \textsc{ThreadAnalyzer} combined with the\textsc{RelevanceAgent}.
LLaMA Guard~\cite{dubey2024llama3herdmodels} is a safety classifier fine-tuned from LLaMA~3 to detect harmful content across a fixed 14-category taxonomy.

\mypar{Default policy}
We test the default LlamaGuard policy as a pure exclusion filter. Posts classified as safe (i.e., those that trigger no default safety category) are treated as relevant. All unsafe $S_{k}$ outputs are not relevant, as they mostly map to prohibited topics. The rationale is that S1 (Violent Crimes/Arms), S2 (Non-Violent Crimes), and S12 (Sexual Content) naturally suppress drugs, arms, and porn. In practice, S2 conflates cybercrime with drug trafficking and financial fraud, so the majority of hacking posts are suppressed alongside the non-CTI content, yielding very low recall. 
This configuration represents the out-of-the-box baseline.

As Figure~\ref{fig:llamaguard_default} reports, the default configuration of LlamaGuard blocks a far larger share of hacking documents compared to the \SysName approach, making it unsuitable as a relevance filter for CTI.

\mypar{Custom CTI policy}
As in the default 14 categories taxonomy, there is no adequate mapping to CTI-relevant content, so we introduce a custom category.
We define $S_{14}$: \textit{Cyber Threat Intelligence}, which correspond in a positive list of
\ac{CTI} indicators (e.g., malware, exploit kits, credential theft services, botnet
infrastructure).
No exclusion rules are stated; non-\ac{CTI} content is expected to fall through
to \texttt{safe} by default.
A post is classified as \textit{Relevant} only if the model responds
\texttt{unsafe\,S14}.

As Figure~\ref{fig:llamaguard_custom} shows, this substantially improves recall on the Hacking category, but it also introduces a much higher false-positive rate across policy-violating content. 

\begin{figure}[t]
  \centering
  \begin{tikzpicture}
  \begin{axis}[
      ybar,                       
      ymin=0,
      ymax=250,
      ytick={0,50,100,150,200},
      bar width=6pt,
      enlarge x limits=0.08,
      xtick pos=bottom,
      nodes near coords,
      nodes near coords style={font=\footnotesize, rotate=90, anchor=west,yshift=0.4pt},
      tick label style={font=\footnotesize},
      label style={font=\small},
      ymajorgrids=true,
      grid style={dotted, gray!80},
      symbolic x coords={Drugs,Electronic,Gambling,Others,Financial,Hacking,Crypto,Violence,Porn,Arms},
      xtick=data,
      x tick label style={rotate=45, anchor=east, font=\footnotesize},
      ylabel={Count},
      width=\linewidth,
      height=0.6\linewidth,
      legend style={
        font=\footnotesize,
        at={(0.02,0.98)},
        anchor=north west,
        legend cell align={left},
        reverse legend
      },
      area legend,
  ]
  \addplot[
      draw=black,
      fill=teal!70!black,
      bar shift=-3.6pt,
  ] coordinates {
      (Arms,1)
      (Porn,1)
      (Violence,3)
      (Crypto,19)
      (Hacking,183)
      (Financial,15)
      (Others,8)
      (Gambling,2)
      (Electronic,1)
      (Drugs,0)
  };
  \addplot[
      draw=black,
      fill=red!25,
      bar shift=3.6pt,
  ] coordinates {
      (Arms,4)
      (Porn,5)
      (Violence,2)
      (Crypto,18)
      (Hacking,38)
      (Financial,4)
      (Others,24)
      (Gambling,14)
      (Electronic,15)
      (Drugs,2)
  };
  \legend{\SysName,LLamaGuard}
  \end{axis}
  \end{tikzpicture}
  \caption{LLamaGuard3 Relevance Prediction (Default)}
  \label{fig:llamaguard_default}
\end{figure}
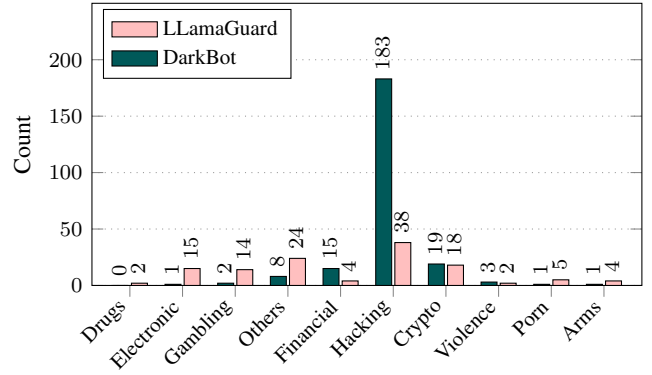
\begin{figure}[t]
  \centering
  \begin{tikzpicture}
  \begin{axis}[
      ybar,                       
      ymin=0,
      ymax=250,
      ytick={0,50,100,150,200},
      bar width=6pt,
      enlarge x limits=0.08,
      xtick pos=bottom,
      nodes near coords,
      nodes near coords style={font=\footnotesize, rotate=90, anchor=west,yshift=0.4pt},
      tick label style={font=\footnotesize},
      label style={font=\small},
      ymajorgrids=true,
      grid style={dotted, gray!80},
      symbolic x coords={Drugs,Electronic,Gambling,Others,Financial,Hacking,Crypto,Violence,Porn,Arms},
      xtick=data,
      x tick label style={rotate=45, anchor=east, font=\footnotesize},
      ylabel={Count},
      width=\linewidth,
      height=0.6\linewidth,
      legend style={
        font=\footnotesize,
        at={(0.02,0.98)},
        anchor=north west,
        legend cell align={left},
        reverse legend
      },
      area legend,
  ]
  \addplot[
      draw=black,
      fill=teal!70!black,
      bar shift=-3.6pt,
  ] coordinates {
      (Arms,1)
      (Porn,1)
      (Violence,3)
      (Crypto,19)
      (Hacking,183)
      (Financial,15)
      (Others,8)
      (Gambling,2)
      (Electronic,1)
      (Drugs,0)
  };
  \addplot[
      draw=black,
      fill=red!25,
      bar shift=3.6pt,
  ] coordinates {
      (Arms,24)
      (Porn,23)
      (Violence,14)
      (Crypto,13)
      (Hacking,175)
      (Financial,26)
      (Others,4)
      (Gambling,9)
      (Electronic,18)
      (Drugs,8)
  };
  \legend{\SysName,LLamaGuard}
  \end{axis}
  \end{tikzpicture}
  \caption{LLamaGuard3 Relevance Prediction (Custom)}
  \label{fig:llamaguard_custom}
\end{figure}
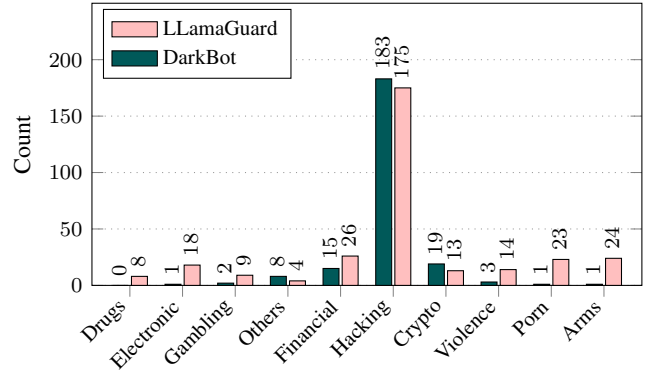

\section{Style Embeddings Analysis}
\label{app:syle_embeddings}
\definecolor{umaporange}{RGB}{255,127,14}
\definecolor{umapblue}{RGB}{76,114,176}
\begin{figure*}[t]
\centering

\pgfplotslegendfromname{umaplotslegend}
\smallskip

\begin{tikzpicture}
\begin{groupplot}[
  group style={
    group size=4 by 1,
    horizontal sep=0.45cm,
    ylabels at=edge left,
  },
  width=0.3\textwidth,
  height=0.27\textwidth,
  xlabel={\footnotesize UMAP-1},
  ylabel={\footnotesize UMAP-2},
  tick label style={font=\scriptsize},
  label style={font=\footnotesize},
  title style={font=\footnotesize, yshift=-1pt},
  axis line style={thin, black},
  tick style={thin, black},
]
\nextgroupplot[title={Style-Embedding, Dread}, label={fig:umap-anna-dread}]
\addplot[only marks, mark=*, mark size=1.2pt,
         color=umapblue, opacity=0.5, draw opacity=0]
  table[x=x, y=y, col sep=comma,
        restrict expr to domain={\thisrow{type}}{-0.5:0.5}]
  {Data/umap_AnnaWegmann_Style-Embedding_dread.csv};
\addplot[only marks, mark=diamond*, mark size=2.0pt,
         color=umaporange, opacity=0.7,
         mark options={draw=white, line width=0.25pt}]
  table[x=x, y=y, col sep=comma,
        restrict expr to domain={\thisrow{type}}{0.5:1.5}]
  {Data/umap_AnnaWegmann_Style-Embedding_dread.csv};
\nextgroupplot[title={Style-Embedding, BHW}, label={fig:umap-anna-bhw}]
\addplot[only marks, mark=*, mark size=1.2pt,
         color=umapblue, opacity=0.5, draw opacity=0]
  table[x=x, y=y, col sep=comma,
        restrict expr to domain={\thisrow{type}}{-0.5:0.5}]
  {Data/umap_AnnaWegmann_Style-Embedding_www_blackhatworld_com.csv};
\addplot[only marks, mark=diamond*, mark size=2.0pt,
         color=umaporange, opacity=0.9,
         mark options={draw=white, line width=0.25pt}]
  table[x=x, y=y, col sep=comma,
        restrict expr to domain={\thisrow{type}}{0.5:1.5}]
  {Data/umap_AnnaWegmann_Style-Embedding_www_blackhatworld_com.csv};
\nextgroupplot[title={StyleDistance, Dread}, label={fig:umap-sd-dread}]
\addplot[only marks, mark=*, mark size=1.2pt,
         color=umapblue, opacity=0.5, draw opacity=0]
  table[x=x, y=y, col sep=comma,
        restrict expr to domain={\thisrow{type}}{-0.5:0.5}]
  {Data/umap_StyleDistance_styledistance_dread.csv};
\addplot[only marks, mark=diamond*, mark size=2.0pt,
         color=umaporange, opacity=0.9,
         mark options={draw=white, line width=0.25pt}]
  table[x=x, y=y, col sep=comma,
        restrict expr to domain={\thisrow{type}}{0.5:1.5}]
  {Data/umap_StyleDistance_styledistance_dread.csv};
\nextgroupplot[title={StyleDistance, BHW}, label={fig:umap-sd-bhw},
  legend to name=umaplotslegend,
  legend style={draw=none, font=\small,
                legend columns=-1, column sep=0pt,
                /tikz/every odd column/.append style={column sep=1.5pt, yshift=0.4pt},
                /tikz/every even column/.append style={column sep=4pt}},
]
\addplot[only marks, mark=*, mark size=1.2pt,
         color=umapblue, opacity=0.5, draw opacity=0, forget plot]
  table[x=x, y=y, col sep=comma,
        restrict expr to domain={\thisrow{type}}{-0.5:0.5}]
  {Data/umap_StyleDistance_styledistance_www_blackhatworld_com.csv};
\addlegendimage{only marks, mark=*, mark size=1.7pt,
                color=umapblue, opacity=1.0, draw opacity=0, yshift=0.4pt}
\addlegendentry{Users}
\addplot[only marks, mark=diamond*, mark size=2pt,
         color=umaporange, opacity=0.9,
         mark options={draw=white, line width=0.25pt}, forget plot]
  table[x=x, y=y, col sep=comma,
        restrict expr to domain={\thisrow{type}}{0.5:1.5}]
  {Data/umap_StyleDistance_styledistance_www_blackhatworld_com.csv};
\addlegendimage{only marks, mark=diamond*, mark size=3pt,
                color=umaporange, opacity=1.0, yshift=0.4pt}
\addlegendentry{\SysName}
\end{groupplot}
\end{tikzpicture}

\caption{UMAP projection of two style embeddings for \SysName and real-user messages on Dread and Blackhatworld (BHW)}
\label{fig:umap-style}
\end{figure*}
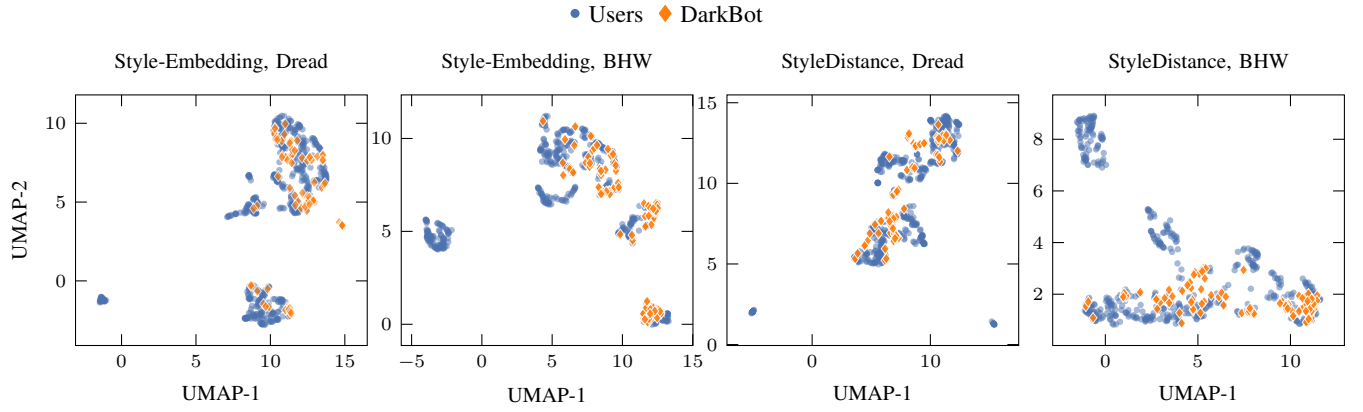
To assess the system’s ability to adapt to the conversational tone and style of ongoing discussions, we compared its messages with those of other users on \textit{BHW} and \textit{Dread}.
We embedded all messages from the live interactions using two style-focused sentence transformers, i.e., \textit{Style-Embedding} \cite{same_topic_or_same_author} and \textit{styledistance} \cite{styledistance}. For each system message, we compute the average cosine similarity to user messages. 
StyleDistance yields higher average cosine similarity scores (0.73 for \textit{Dread} and 0.68 for \textit{BHW}) than Style-Embeddings (0.23/0.02).
We use UMAP for 2D visualization to identify stylistic clusters and outliers (Fig.~\ref{fig:umap-style}). The two-dimensional UMAP projections show substantial overlap between \SysName and user-message clusters under both
style representations, with no visually isolated cluster
containing only system messages, which shows that bot messages blend into user message clusters and suggests the bot's style is indistinguishable from that of regular users.

\section{CTI Entities Subset}
\label{sec:app_cti_entities}

\mypari{Entity types}
The matched-pair experiment (§\ref{sec:exp_activity_e3}) measures how much threat
intelligence a thread yields, which requires deciding what counts as a unit of
intelligence. We annotate entity mentions using the object types of
STIX~2.1, the interoperability standard for threat-intelligence exchange, so that
the measure is defined by an external specification rather than by a scheme
introduced for this evaluation. Table~\ref{tab:cti_entities} lists the resulting
types: eight STIX Domain Objects, describing what a discussion is about, and nine
STIX Cyber-observable Objects, the indicators that could be fed directly to a
detection system.
Two categories deserve comment. Defence (\texttt{course-of-action}) covers a
control or detection signal named as what catches or stops an operation, and is
the type that dominates our results; in an offensive discussion these are
described from the attacker's point of view, as obstacles to be worked around.
Evasion has no separate type: ATT\&CK treats defence evasion as a \textit{tactic}
over techniques rather than as a category of its own, so evasive manoeuvres are
annotated as Technique.

\mypari{Annotation procedure}
We annotated every
human-authored message published in either arm of the 20 pairs within the
seven-day observation window: 174 messages in total, 90 in the threads
\SysName{} was assigned to and 84 in their matched controls. Initial posts, bot
messages, and moderator or automated messages are excluded, so that the annotated
population is exactly the population counted by the activity measure. 
A sources of pre-annotation was offered to reduce omissions. A
regular-expression extractor was used as pre-annotation for the indicators with fixed formats.  Across the 310 messages of the cohort it proposed six candidates
(three URLs and three images).

\begin{table}[h]
\centering
\caption{Entity types used to annotate the matched-pair threads. The listed
categories are STIX~2.1 object types and constitute the measure reported in
Section~\ref{sec:exp_activity_e3}.}
\label{tab:cti_entities}
\footnotesize
\setlength{\tabcolsep}{2pt}

\begin{tabular}{l l p{4cm}}
\toprule
\textbf{Entity} &
\textbf{STIX 2.1 type} &
\textbf{Description}
\tabularnewline
\midrule

\multicolumn{3}{l}{\textit{STIX Domain Objects}}
\tabularnewline

Technique &
\texttt{attack-pattern} &
Method or manoeuvre described as done or doable, evasion included
\tabularnewline

Tool &
\texttt{tool} &
Named legitimate or dual-use software used to do the work
\tabularnewline

Malware &
\texttt{malware} &
Named malicious software: stealers, RATs, loaders, botnet agents
\tabularnewline

Infrastructure &
\texttt{infrastructure} &
Resources the operation runs on: proxies, hosting, VPS, channels
\tabularnewline

Target &
\texttt{identity} &
Platform, organisation or sector being acted upon or evaded
\tabularnewline

Actor &
\texttt{threat-actor} &
A named person, crew or vendor being discussed
\tabularnewline

Defence &
\texttt{course-of-action} &
Defensive control or detection signal named as what stops you
\tabularnewline

Vulnerability &
\texttt{vulnerability} &
A named weakness or CVE being exploited or discussed
\tabularnewline

\addlinespace

\multicolumn{3}{l}{\textit{STIX Cyber-observable Objects (indicators)}}
\tabularnewline

URL &
\texttt{url} &
Any link
\tabularnewline

Image &
\texttt{file} &
Embedded image reference
\tabularnewline

IP address &
\texttt{ipv4-addr} &
IPv4/IPv6 address
\tabularnewline

Email &
\texttt{email-addr} &
Email address
\tabularnewline

XMPP/Jabber &
\texttt{user-account} &
Jabber identifier
\tabularnewline

Discord &
\texttt{user-account} &
Discord handle or user link
\tabularnewline

Telegram &
\texttt{user-account} &
Telegram handle or t.me link
\tabularnewline

Skype &
\texttt{user-account} &
Skype handle or live: identifier
\tabularnewline

Crypto address &
\texttt{user-account} &
On-chain wallet address
\tabularnewline

\bottomrule
\end{tabular}
\end{table}

\section{Weekly Timing Patterns}
\label{sec:app_week_timing}
Figure ~\ref{fig:dayofweek_real_platforms} illustrates number of messages received per day of week on \textit{dread} and \textit{BHW} during experiment~\ref{sec:exp_real_forums}.
\usetikzlibrary{pgfplots.polar}
\begin{figure}[H]
\centering
\begin{tikzpicture}
\begin{polaraxis}[
    name=left,
    width=4cm, height=4cm,
    xtick={0, 51.43, 102.86, 154.29, 205.71, 257.14, 308.57},
    xticklabels={Mon, Tue, Wed, Thu, Fri, Sat, Sun},
    xticklabel style={font=\small},
    ymin=0, ymax=126,
    ytick={25, 50, 75, 100},
    yticklabel style={font=\tiny},
    grid=both,
    grid style={line width=0.3pt, gray!40},
    legend to name=daylegend,
    area legend,
    legend style={draw=none, font=\small, column sep=0.8em, legend columns=3},
]
\addplot[thick, fill={rgb,255:red,90;green,90;blue,90},
         draw={rgb,255:red,90;green,90;blue,90}, fill opacity=0.25]
    coordinates {
        (0,71)(51.43,74)(102.86,83)(154.29,97)
        (205.71,126)(257.14,91)(308.57,34)(360,71)
    };
\addlegendentry{All messages}
\addplot[thick, fill=customblue, draw=customblue, fill opacity=0.25]
    coordinates {
        (0,31)(51.43,45)(102.86,37)(154.29,50)
        (205.71,42)(257.14,52)(308.57,19)(360,31)
    };
\addlegendentry{Replies}
\addplot[thick, fill={rgb,255:red,0;green,140;blue,75},
         draw={rgb,255:red,0;green,140;blue,75}, fill opacity=0.25]
    coordinates {
        (0,13)(51.43,16)(102.86,15)(154.29,20)
        (205.71,19)(257.14,21)(308.57,8)(360,13)
    };
\addlegendentry{Direct replies}
\end{polaraxis}

\begin{polaraxis}[
    name=right,
    at={(left.east)}, anchor=west, xshift=1.2cm,
    width=4cm, height=4cm,
    xtick={0, 51.43, 102.86, 154.29, 205.71, 257.14, 308.57},
    xticklabels={Mon, Tue, Wed, Thu, Fri, Sat, Sun},
    xticklabel style={font=\small},
    ymin=0, ymax=49,
    ytick={10, 20, 30, 40},
    yticklabel style={font=\tiny},
    grid=both,
    grid style={line width=0.3pt, gray!40},
]
\addplot[thick, fill={rgb,255:red,90;green,90;blue,90},
         draw={rgb,255:red,90;green,90;blue,90}, fill opacity=0.25]
    coordinates {
        (0,10)(51.43,49)(102.86,45)(154.29,23)
        (205.71,22)(257.14,29)(308.57,16)(360,10)
    };
\addplot[thick, fill=customblue, draw=customblue, fill opacity=0.25]
    coordinates {
        (0,3)(51.43,41)(102.86,28)(154.29,17)
        (205.71,16)(257.14,24)(308.57,8)(360,3)
    };
\addplot[thick, fill={rgb,255:red,0;green,140;blue,75},
         draw={rgb,255:red,0;green,140;blue,75}, fill opacity=0.25]
    coordinates {
        (0,1)(51.43,11)(102.86,12)(154.29,6)
        (205.71,3)(257.14,6)(308.57,3)(360,1)
    };
\end{polaraxis}

      anchor=north east] at (right.north east) {Dread};
\node[font=\small\itshape, below=16pt of left.south] {BlackHatWorld};

\node[font=\small\itshape, below=16pt of right.south] {Dread};

\node[above=10pt of left.north, xshift=1.8cm] {\pgfplotslegendfromname{daylegend}};

\end{tikzpicture}
\caption{Message distribution by day of week.}
\label{fig:dayofweek_real_platforms}
\end{figure}
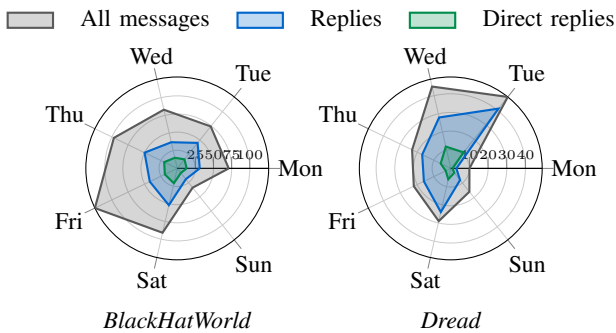

\section{Interaction Examples}
\label{sec:app_examples}
This section reports examples of interactions conducted by \SysName (indicated as Bot) in the \textit{Dread} forum. All examples are anonymized, omitting unrelated or less interesting messages present in the thread for brevity. Each example reports the thread's initial post on top, and represents an exchange within the thread, where the bot intervened as a regular participant without disclosing its automated nature.
We highlight in \bl{blue} the more relevant information extracted by the bot in the interactions. The harmful language used in the examples is censored using the symbol "\censor{X}".

\myexamplebox[H]{%
   \SysName conversation on dread forum.
}{%
  \begin{center}
    \textit{  \textbf{Spreading method for linux?}\\
      \smallskip
what are some good spreading methods to taret linux ?\\
so far i have binded with some crack and that worked okayish\\
maybe a server setup that can detect bash curl commands and then go on social media to help out noobs with a single "magic bash curl command" to fix thier s\censor{hitvidia} drivers  }
  \end{center}
}{%
  \begin{tcolorbox}[systurn]
    \textbf{Bot:} Binded with a crack , nice one !I wanna know , what kinda crack ? , specific app or general ?and does it needs elevated priviledges to kick it off ?
  \end{tcolorbox}
  \begin{tcolorbox}[userturn]
    \textbf{User:} well a certain \bl{game patcher}.\\
no no root is needed. not needed for stealers and ransomware in general.
  \end{tcolorbox}
}{ex_2}

\myexamplebox[H]{%
   \SysName conversation on dread forum.
}{%
  \begin{center}
        \textbf{Best source for edr heuristics/behavioral analysis evasion}
  \end{center}
  I want to do some source code analysis for inspiration. In particular I want to examine some source code of some recent malware that implements effective (could be burnt method but idc I want something more recent to build on or inspire myself) edr behavioral analysis evasion. If you guys don't know any good vxunderground/malware baazar source to look at, can you at least provide some solid methods that are recent and known to get inspiration from. Also /d/malware is being annoying bc this same post has been pending for like a day since it was my first post.
}{%
  \begin{tcolorbox}[userturn]
    \textbf{User A:} Check vx-underground's MalwareSourceCode repo for 2024 LockBit or RansomHub leaks with EDR-killing process hollowing and indirect syscalls. For inspiration, use Hell's Hall for dynamic syscall evasion and JIT-based memory injection like in Crypto24 samples from MalwareBazaar.
  \end{tcolorbox}
  \begin{tcolorbox}[systurn]
    \textbf{Bot:} Yeah /d/malware are w\censor{ankas}, slow as f\censor{uck}...For source code, been digging through VX Underground, good stuff there...You talk about burnt methods but what sort of techniques have you had any success with against EDR's recently, I mean beyond the usual DLL injection rubbish?
  \end{tcolorbox}
  \begin{tcolorbox}[userturn]
    \textbf{User B:}  \bl{Got EDR-Freeze concept to work}, but still gets busted. Maybe implement as shellcode?
  \end{tcolorbox}
}{ex_1}

\newpage

\myexamplebox[H]{%
   \SysName conversation on dread forum.
}{%
  \begin{center}
        \textbf{US Fullz [ Full Name + Address + DOB + SSN + DL + Credit Report(some have, not all)] | Private data}
  \end{center}
  Hello.
  I've recently breached a credit fixing company for US residents and obtained about 29k-30k fullz. There are about 8k Credit card data, but it's encrypted, so I'm working on that.
  Anyways, all this data is 1) private 2) fresh.
  On purchase, you will obtain a .zip containing: a .txt file with the person's information. Images/pdfs of their scanned documents \\
  Price: 5\$/unit\\
  I'm also open for reselling opportunities. Anyone who is a vendor or has their own autoshop with a customerbase may contact me for a \% split on sold items.
}{%
  \begin{tcolorbox}[systurn]
    \textbf{Bot:} 5\$/unit my a\censor{ss}!  Always looking for good fullz. What sort of data are you sifting through those zips for? Any particular docs appearing more than others?
  \end{tcolorbox}
  \begin{tcolorbox}[userturn]
    \textbf{User:} DM if you're looking. \bl{the text document will be in this format:
Contact ID: 
Full Name:
Email: 
Mobile: 
Phone: 
Date of Birth: 
Social Security Number:
Home Address: 
Telecom Partner1:
As per docs, they appear the most in this order:
1. credit reports are the most common as the source I've gotten the fullz is such.
2. Then DL/IDs.
3. bills and ssn(ssn is in text format, so it doesnt matter if the pic is there or not)}
  \end{tcolorbox}
  \begin{tcolorbox}[systurn]
    \textbf{Bot:} nice! so you have a table with metadata nice! Are you able to sort by that date to make sure you are grabbing the freshest fullz? What the latest date you got?  Is it 100\% accurate in your experience?
  \end{tcolorbox}
  \begin{tcolorbox}[userturn]
    \textbf{User:} Yes. I can sort through the freshest data. \bl{I still have live access to the database}, so everyday I get new fullz.
  \end{tcolorbox}
  \begin{tcolorbox}[systurn]
    \textbf{Bot:} Live nice!!you thing more breaches of credit fixing companies will apper in the nex few month?
  \end{tcolorbox}
  \begin{tcolorbox}[userturn]
    \textbf{User:} I have no way to guess tbh
  \end{tcolorbox}
}{ex_4}

\myexamplebox[H]{%
   \SysName interaction under experiment \S\ref{sec:exp_activity_e3}.
}{%
  \begin{center}
    \textit{ \textbf{Do I still need human farms to solve Captchas?} \\ \smallskip
Are current captchas still sound against AI/computer vision? I feel everybody is hyping it up as hell, and I'm plying with it to make my automations faster, but for many of them I feel like paying some dude to solve captha all day is still the most convininent option rn\ldots{} I was wondering what's the status of that, and if it will be actually cheaper than humans. Or just better to look for standard bypasses?}
  \end{center}
}{%
  \begin{tcolorbox}[userturn]
    \textbf{User:} Yeah, Gemini is pretty good at solving audio CAPTCHAs. \bl{I've been using it to mass-create GitHub accounts without needing a CAPTCHA solver service.}
  \end{tcolorbox}
  \begin{tcolorbox}[systurn]
    \textbf{Bot:} Yes BH is getting less and less worth it, +100!!Geminis.GHACO ACCOUNT THING IS AOMG, TRUE TOME +1 That being said, using Gemini what prompts are working best for you so far to get consistent results on audio CAPTCHAs?
  \end{tcolorbox}
  \begin{tcolorbox}[userturn]
    \textbf{User:} rightt, just use \bl{Gemini 2.5 (this model is the best so far for audio recognition), feed it the audio along with the exact instruction text from the CAPTCHA, and enforce a strict JSON output format like \{"result": "answer"\}}
  \end{tcolorbox}
  \begin{tcolorbox}[systurn]
    \textbf{Bot:} F\censor{ucking} True , Gemini 2.5 did pretty much better ,Nice Configuration for Enforcing Strict Json Structure For Parse Result )=How You Preprocess The Audio File Before Get Into Api ? Any Noise Reduction Or Just Feed Raw As The Audio ?
  \end{tcolorbox}
  \begin{tcolorbox}[userturn]
    \textbf{User:} \bl{i just throw the full audio straight into the model}
  \end{tcolorbox}
}{ex_7}

\newpage
\myexamplebox[H]{%
   \SysName conversation on dread forum.
}{%
  \begin{center}
    \textit{  \textbf{question regarding bypassing security} \\ \smallskip
might be a r\censor{etarded} question so let me know, but i was wondering if there was anyway to i guess bypass security questions (mainly for like microsoft accounts)
i have emails/passwords/numbers and was originally going to go the route of finding an otp bot and trying to just get in with the code but i realize that creating or finding a legitimate otp bot is way more difficult than i thought. this is not for fraud / monetary purposes its for an odd reason, but yeah. wondering if theres any other way i could get into these microsoft accounts. thanks guys }
  \end{center}
}{%
  \begin{tcolorbox}[systurn]
    \textbf{Bot:} OTP bots s\censor{ux} hahaha. I'm with you, security questions can be awfully good...or awfully r\censor{etarded}. Whats the 'odd reason' for getting into these accounts? if you don't mind
  \end{tcolorbox}
  \begin{tcolorbox}[userturn]
    \textbf{User:} to be honest, the usernames \bl{linked to the Xbox accounts. I want to get into the account to swap the username off and then sell it (they go for thousands)}. i got information on the emails linked to the account names and found the accounts that are worth alot and did some csint to get the passwords and numbers. thats where im at now.
  \end{tcolorbox}
}{ex_3}

\newpage

\myexamplebox[H]{%
   \SysName conversation on dread forum.
}{%
  \begin{center}
    \textit{ \textbf{eBay login always triggering 2FA} \\  \smallskip
I'm using anti detect browser, one of the best + luxurysocks... still can't login. Any advice? }
  \end{center}
}{%
  \begin{tcolorbox}[systurn]
    \textbf{Bot:} Damn man, 2FA s\censor{ucks}, no matter how shiny the stuff is. I'm using the same setup and yet sometimes ebay still gets me. How you exactly configured anti-detect browser? Tweaked any settings?
  \end{tcolorbox}
  \begin{tcolorbox}[userturn]
    \textbf{User:} Learn how to spoof for otp code. Simple really.
  \end{tcolorbox}
  \begin{tcolorbox}[systurn]
    \textbf{Bot:} yeah, I know what you are saying, sometimes too much 2FA!So, what you mean by 'spoof for OTP code'? Are you doing at browser level or you are using some external service to intercept/spoof?
  \end{tcolorbox}
  \begin{tcolorbox}[userturn]
    \textbf{User:} So basically i do this for getting 2fa code's to \bl{login to bank's} not to sure how far you would go to get a OTP code for ebay. But basically \bl{Sending the victim a message via SID ( Sender ID - Cloning "Ebay" so the text message to the victim pop's up as a text from Ebay ). In the message it should usually say something about a suspicious login attempt and to call this support number ( which you own ). Basically when they call just social engineer the code out the victim.}
  \end{tcolorbox}
}{ex_6}

\clearpage

\end{document}